\documentclass[aps,pra,reprint,twocolumn,showpacs,showkeys,floatfix,nobibnotes]{revtex4-2}
\usepackage[utf8]{inputenc}
\usepackage{amsfonts}
\usepackage{multirow}
\usepackage{float}
\usepackage{caption,subcaption}
\usepackage{bm}
\usepackage{amssymb,amsmath}
\usepackage{graphicx,mathptmx,physics}
\usepackage{epstopdf}
\usepackage[
colorlinks=true,
citecolor=blue,
linkcolor=blue,
urlcolor=blue
]{hyperref}
\allowdisplaybreaks 
\begin{document}
	
	\title{Quantum teleportation using photon-added two-mode squeezed Fock states: A comparison with photon-subtracted counterparts}
	

	
	

	\author{Ankita}
	\author{Arpita Chatterjee}
	\email{Corresponding author: arpita.sps@gmail.com}
	
	\affiliation{Department of Mathematics, J. C. Bose University of Science and Technology, YMCA, Faridabad 121006, Haryana, India}
		%
	\begin{abstract}
		Quantum state engineering frequently employs non-Gaussian operations, particularly photon addition and subtraction, to tailor and improve the nonclassical characteristics of quantum states. We investigate how photon addition to a two-mode squeezed Fock state (TMSFS) influences its performance in the Braunstein-Kimble quantum teleportation protocol. We obtain the Wigner characteristic function of the photon-added TMSFS (PA-TMSFS) and evaluate the corresponding teleportation fidelities for both Gaussian and non-Gaussian input states. 
		We further compare the teleportation performance of PA-TMSFS with that of its photon-subtracted counterpart by accessing their fidelity enhancements over the parent TMSFS and considering the probabilistic nature of both the operations.
		Our results demonstrate that photon subtraction provides superior teleportation performance compared with photon addition, with this advantage becoming more pronounced when the success probabilities of the corresponding non-Gaussian operations are considered. In particular, the symmetric $(1,1)$ photon subtraction yields the optimal trade-off between teleportation fidelity and success probability among the configurations investigated. These findings highlight that photon subtraction retains its advantage 
		even when the underlying entangled resource is already non-Gaussian.
		
		
	\end{abstract}
	\date{\today}	
	\maketitle
	\newpage
	\section{Introduction}	
	Quantum technologies \cite{Ekert1991, Steane1998, Giovannetti2006, Pirandola2018, Pirandola2020, Wolf2021} have witnessed growing research interest because of their potential to accomplish information processing and communication tasks \cite{Bennett1992, Bollinger1996, Cirac1999, Hillery1999, Pati2000, Shor2000, Knill2001, Fitzi2001, Menicucci2006, Giovannetti2008, Pironio2010, Ray2016, Das2021, Singh2024} that extend beyond the limits of classical approaches \cite{Hardy1999,Harrow2017,Kumar2019,Heinosaari2024}. Quantum teleportation (QT) \cite{Horodecki1996,Pirandola2015,Sohail2023,Hu2023,DAurelio2025} represents one of the fundamental protocols of quantum information science, enabling the transfer of
	quantum information between spatially separated systems without physically transmitting the quantum state itself. In particular, QT plays a significant role in distributed quantum information processing and is therefore expected to be a key component in the realization of quantum networks and the future quantum internet \cite{Yonezawa2004,Kimble2008}.	
	QT was originally proposed by Bennett \textit{et al.} \cite{Bennett1993} for discrete-variable systems and was subsequently extended to continuous-variable (CV) systems. In the CV teleportation protocol, an unknown quantum state is transferred from a sender (Alice) and reconstructed at a remote receiver's end (Bob) using a shared entangled resource and a classical communication.
	The quality of the teleportation is governed by the quantum correlations present in the shared resource as weaker or imperfect correlations generally resulting in less accurate state reconstruction \cite{Adesso2005,DellAnno2010}. The reconstructed state is typically quantified by the teleportation fidelity.

	The two-mode squeezed vacuum (TMSV) state serves as the 
	standard Gaussian entangled resource for CV quantum teleportation \cite{Braunstein1998,Furusawa1998,Dell2007}. 
	However, the finite degree of squeezing available in practical TMSV states  \cite{Braunstein2005} restricts the teleportation fidelity as perfect fidelity can be achievable only in the limit of infinite squeezing \cite{Luiz2015}.
	This limitation has stimulated considerable interest in non-Gaussian quantum states and operations \cite{Cochrane2002, Wenger2004,Dell2007, DellAnno2010, LeeJ2013, Bartley2015}, which offer additional flexibility to manipulate quantum correlations using finite resources.
	Among the various non-Gaussian resources, TMSFS  \cite{KOrolev2024,Bashmakova2025} provides a natural
	extension of the original TMSV channel. The Fock-state excitation introduces a nontrivial photon-number structure and intrinsic non-Gaussian features into the TMSV resource, potentially exhibiting teleportation characteristics distinct from those of conventional Gaussian resources. Accordingly, we investigate the potential of TMSFS for QT and explore how its performance can be tailored through further non-Gaussian engineering.
	
	Photon subtraction (PS) \cite{Kitagawa2006,Zhang2010,Bartley2013,LeeS2013} and photon addition (PA) \cite{Zhang2011,Park2012,Zhang2013,LeeJ2013,Navarrete2012} are widely used operations for engineering optical quantum states. Although both operations modify the photon-number distribution and generate non-Gaussian states, they act differently on the underlying quantum state and therefore lead to distinct performance in quantum information protocols. For Gaussian engineered resources, several studies have reported an advantage of photon subtraction over photon addition for specific applications and performance measures \cite{DellAnno2010,Lee2011,LeeJ2013,Navarrete2012,Chandan2023,Ankita2026}. It is therefore worthwhile to investigate whether this relative advantage is retained when the underlying entangled resource is intrinsically non-Gaussian. This question is nontrivial because the effects of photon addition or subtraction on a TMSFS are governed not only by the nature of the operation itself but also by the preliminary photon-number structure and non-Gaussian correlations of the resource. 
	Moreover, the probabilistic nature of these operations introduces an additional consideration, as their success probabilities affect their practical utility for quantum information processing \cite{Zhang2013,Kumar2022}. In our previous work, we focused on the application of PS-TMSFS to CV quantum teleportation \cite{Ankita2026PLA}. The present study extends this investigation to PA-TMSFS, thereby facilitating a direct comparison of the two operations for an intrinsically non-Gaussian resource. 
	
	Our earlier work \cite{Ankita2026PLA} highlighted that PS-TMSFS can enhance the teleportation fidelity for Gaussian inputs, particularly for weak squeezing and high beam-splitter transmissivity. The symmetric photon subtraction configurations, especially the $(2,2)$ pattern, provided the maximum enhancement. For non-Gaussian inputs, the fidelity enhancement follows a similar trend, although stronger squeezing was generally required to exceed the no-cloning limit. At higher squeezing, the advantage of photon subtraction with respect to the parent TMSV became less pronounced. In this work, we investigate CV quantum teleportation using PA-TMSFS as an entangled resource. We present the scheme for generating PA-TMSFS and analytically derive the corresponding Wigner characteristic function and success probability. The teleportation fidelity is then evaluated for Gaussian and non-Gaussian input states, considering different photon-addition configurations and examining the effects of the squeezing parameter and beam-splitter transmissivity. A comparative analysis of PA-TMSFS and PS-TMSFS is also performed, focusing on fidelity enhancement and probability-weighted enhancement to evaluate their relative effectiveness and associated success probabilities. 
	
	The paper is organized as follows. In section~\ref{WCF}, we describe the preparation of the PA-TMSFS and compute its Wigner characteristic function and success probability. In Section~\ref{fidelity}, we examine the teleportation fidelity for Gaussian and non-Gaussian input states. We compare the photon addition and photon subtraction processes in terms of fidelity enhancement and probability-weighted enhancement in Section~\ref{DelF}. In Section~\ref{conclusion}, we summarize the main findings of our study. Appendix~\ref{App} provides the Wigner characteristic functions of the input states together with the matrices and coefficients appearing in the Wigner function of PA-TMSFS, its success probability, and the teleportation fidelity.	
	\section{PA-TMSFS: Model and Characteristic Function}
	\label{WCF}
\begin{figure}[htbp]
		\centering
		\includegraphics[trim=0pt 0pt 0pt 0pt,clip,width=0.5\textwidth]{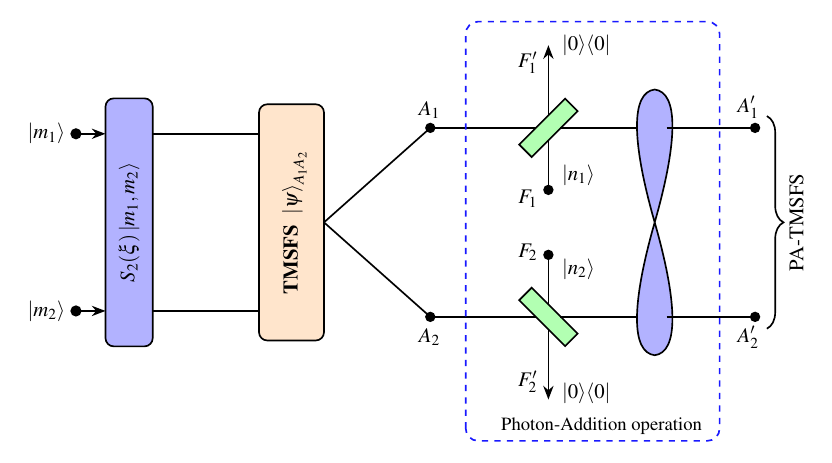}
		\caption{Schematic illustration for generating a photon-added two-mode squeezed Fock state. The initial Fock-state inputs $\ket{m_1}$ and $\ket{m_2}$ are first transformed into a two-mode squeezed Fock state by using the two-mode squeezing operator $S_2(\xi)$. The modes $A_1$ and $A_2$ of the TMSFS are then mixed with ancillary modes $F_1$ and $F_2$ prepared in Fock states $\ket{n_1}$ and $\ket{n_2}$ through beam-splitters with transmissivities $T_1$ and $T_2$, respectively. On-off detectors are applied to the corresponding output modes $F'_1$ and $F'_2$, respectively. Simultaneous detection of the `off' outcomes of both detectors heralds the successful addition of $n_1$ and $n_2$ photons to modes $A_1$ and $A_2$, thereby generating the PA-TMSFS.}
		\label{Fig1}
	\end{figure}

	In this study, the TMSFS acts as the parent state for photon addition, generated by applying the two-mode squeezing operator to an initial two-mode Fock state. Photon addition is then implemented via beam-splitter-based heralding to produce the PA-TMSFS \cite{Dakna1998} as illustrated in Fig.~\ref{Fig1}.
	The derivation of the TMSFS has already been discussed in detail in \cite{Ankita2026PLA}, therefore we omit the corresponding intermediate calculations and use the resulting Wigner characteristic function directly,
	\begin{align}
		\label{WCF_TMSFS}
		&\chi_{A_1,A_2}(\Lambda)
		= \exp\!\Bigg[
		-\frac{1}{4}\Big(
		\cosh(2\xi)(\tau_1^2+\sigma_1^2+\tau_2^2+\sigma_2^2)
		\nonumber\\
		&\qquad\qquad
		-2\sinh(2\xi)(\tau_1\tau_2-\sigma_1\sigma_2)
		\Big)\Bigg]
		\nonumber\\[4pt]
		&\times
		\frac{1}{2^{m_1}m_1!}
		\frac{\partial^{m_1}}{\partial u_1^{m_1}}
		\frac{\partial^{m_1}}{\partial v_1^{m_1}}
		\Bigg[
		2u_1v_1
		+u_1\Big(
		(\cosh \xi\,\tau_1-\sinh \xi\,\tau_2)
		\nonumber\\
		&\qquad
		+i(\cosh \xi\,\sigma_1+\sinh \xi\,\sigma_2)
		\Big)
		-v_1\Big(
		(\cosh \xi\,\tau_1-\sinh \xi\,\tau_2)
		\nonumber\\
		&\qquad
		-i(\cosh \xi\,\sigma_1+\sinh \xi\,\sigma_2)
		\Big)
		\Bigg]_{
			u_1=v_1=0 
		}
		\nonumber\\[6pt]
		&\times
		\frac{1}{2^{m_2}m_2!}
		\frac{\partial^{m_2}}{\partial u_2^{m_2}}
		\frac{\partial^{m_2}}{\partial v_2^{m_2}}
		\Bigg[
		2u_2v_2
		+u_2\Big(
		(-\sinh \xi\,\tau_1+\cosh \xi\,\tau_2)
		\nonumber\\
		&\qquad
		+i(\sinh \xi\,\sigma_1+\cosh \xi\,\sigma_2)
		\Big)
		-v_2\Big(
		(-\sinh \xi\,\tau_1+\cosh \xi\,\tau_2)
		\nonumber\\
		&\qquad
		-i(\sinh \xi\,\sigma_1+\cosh \xi\,\sigma_2)
		\Big)
		\Bigg]_{
			u_2=v_2=0 
		},
	\end{align}
	where $\Lambda=(\tau_1,\sigma_1,\tau_2,\sigma_2)^T$ and $(\tau_j,\sigma_j)$ denotes the phase-space variables of the mode $A_j$ $(j=1,2)$. The photon-added TMSFS is obtained by applying the photon-addition operation to both modes, implemented conditionally through the ancillary modes $F_1$ and $F_2$ and their corresponding beam-splitter interactions.
	Before the beam-splitter operations, the Wigner characteristic function (WCF) of the four mode system can be written as
	\begin{equation}
		\chi_{F_1A_1A_2F_2}(\Lambda) =
		\chi_{A_1A_2}(\Lambda)~
		\chi_{\ket{n_1}}(\Lambda_3)~\chi_{\ket{n_2}}(\Lambda_4), 
	\end{equation}
	where $\chi_{\ket{n_j}}(\Lambda_i), (i=3, 4)$ denotes the Wigner characteristic function of the ancillary Fock state $\ket{n_j}$. After the beam-splitter transformation, it evolves as
	\begin{equation}
		\chi_{F'_1A'_1A'_2F'_2}(\Lambda) =
		\chi_{F_1A_1A_2F_2}(B(T_1, T_2)^{-1}\Lambda). 
	\end{equation}
	The modes $A_1'$ and $A_2'$ represent the output modes of PA-TMSFS obtained after the beam-splitter interaction with the ancillary modes $F_1$ and $F_2$, respectively. The corresponding ancillary output modes $F_1'$ and $F_2'$ are monitored using on-off detectors. The simultaneous detection of the `off' outcomes heralds the successful addition of $n_1$ and $n_2$ photons to modes $A_1$ and $A_2$, respectively. When an equal number of photons is added into each mode, i.e., $n_1 = n_2 = n$, the operation is symmetric, yielding a symmetric PA-TMSFS, whereas $n_1\neq n_2$ corresponds to asymmetric photon addition and results in an asymmetric PA-TMSFS. After the conditional photon addition operation and projection onto the required ancillary states, the unnormalized WCF of the resulting PA-TMSFS can be written as
	\begin{align} 
		\chi_{A'_1A'_2}^{\sim PA} (\Lambda)
		=&
		\frac{1}{(2\pi)^2}
		\int d^2 \Lambda_3 \, d^2 \Lambda_4 \;
		\underbrace{\chi_{F_1' A_1' A_2' F_2'}(\Lambda)}_{\text{joint four-mode characteristic function}}\nonumber \\ & \qquad
		\underbrace{\chi_{|0\rangle}(\Lambda_3)}_{\text {vacuum state projection}} \quad
		\underbrace{\chi_{|0\rangle}(\Lambda_4)}_{\text{ vacuum state projection}} \label{Int}
	\end{align}
	Here $\Lambda_3$ and $\Lambda_4$ denote the phase-space variables correspond to the ancillary modes $F'_1$ and $F'_2$, respectively.
	Integration of Eq.~\eqref{Int} yields
	\begin{align}
		\widetilde{\chi}^{\mathrm{PA}}_{A'_1A'_2}(\Lambda)
		=
		\frac{\hat F}{a_0}
		\exp\left(
		\Lambda^{T}M_1\Lambda
		+u^{T}M_2\Lambda
		+u^{T}M_3u
		\right),
		\label{PAWCF}
	\end{align}
	where $u=(u_1,v_1,u_2,v_2,u'_1,v'_1,u'_2,v'_2)^T$, and the matrices $M_1$, $M_2$ and $M_3$ are given by Eqs.~\eqref{M_1}, \eqref{M_2} and \eqref{M_3} of the Appendix \ref{App}. Further, the differential operator $\hat{F}$ is defined as
	\begin{align*}
		\hat{F}  = &\;
		\frac{2^{-(m_1+m_2+n_1+n_2)}}{m_1!m_2!n_1!n_2!}
		\;
		\frac{\partial^{m_1}}{\partial u_1^{m_1}}
		\;
		\frac{\partial^{m_1}}{\partial v_1^{m_1}}
		\;
		\frac{\partial^{m_2}}{\partial u_2^{m_2}}
		\;
		\frac{\partial^{m_2}}{\partial v_2^{m_2}} \\[6pt]
		&\quad
		\frac{\partial^{n_1}}{\partial {u'_1}^{n_1}}
		\;
		\frac{\partial^{n_1}}{\partial {v'_1}^{n_1}}
		\;
		\frac{\partial^{n_2}}{\partial {u'_2}^{n_2}}
		\;
		\frac{\partial^{n_2}}{\partial {v'_2}^{n_2}}
		\Bigg|_{\substack{
				u_1=v_1=u_2=v_2=0 \\
				u'_1=v'_1=u'_2=v'_2=0
		}}
	\end{align*}
	The normalized WCF corresponding to Eq.~\eqref{PAWCF} is given by
	\begin{equation}
		\chi^{\mathrm{PA}}_{A'_1A'_2}(\Lambda)
		=
		(P^{\mathrm{PA}})^{-1}
		\widetilde{\chi}^{\mathrm{PA}}_{A'_1A'_2}(\Lambda),
		\label{normalizedPA}
	\end{equation}
	where $P^{\mathrm{PA}}$  is the normalization constant associated with the success probability of the photon addition operation. 
	\subsection{Success probability of photon addition}
	The success probability $P^{\mathrm{PA}}$ can be obtained by evaluating the normalized WCF at the origin of phase space i.e. $\chi^{\mathrm{PA}}_{A'_1A'_2}(0)=1$. That means
	\begin{equation}
		\begin{aligned}
			P^{\mathrm{PA}}
			&	=
			\left.
			\widetilde{\chi}^{\mathrm{PA}}_{A'_1A'_2}(\Lambda)
			\right|_{\tau_1=\sigma_1=\tau_2=\sigma_2=0}
			=
			\frac{\hat{F}}{a_0}
			\exp\left(u^{T}M_3u\right).
		\end{aligned}
	\end{equation}
	The success probability therefore depends on the squeezing parameter, beam-splitter transmissivities, and photon numbers associated with the photon addition process. These dependencies are investigated to identify the parameter regimes that allow photon addition with a reasonably high success probability.
	\begin{figure}[htbp]
		\centering
		\includegraphics[trim=0pt 0pt 0pt 0pt,clip,width=0.5\textwidth]{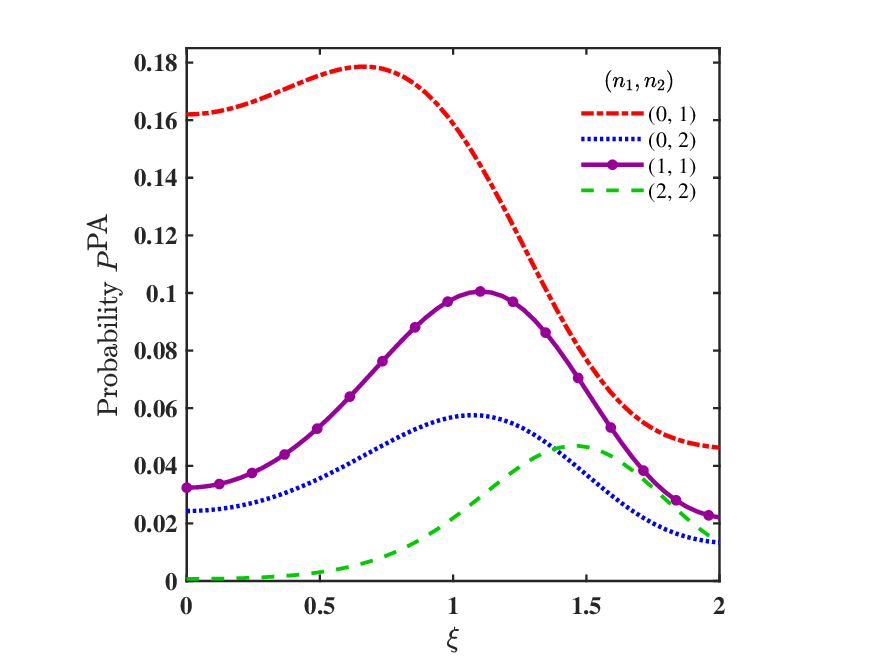}
		\caption{Success probability $P^{\mathrm{PA}}$ as a function of the squeezing parameter $\xi$ for different photon addition configurations $(n_1,n_2)$ of the PA-TMSFS channel at fixed transmissivity $T_1=T_2=0.9$ and initial Fock number $(m_1,m_2)=(1,1)$.}
		\label{Prob_r}
	\end{figure}
	
	\begin{figure}[htbp]
		\centering
		\includegraphics[trim=0pt 0pt 0pt 0pt,clip,width=0.48\textwidth]{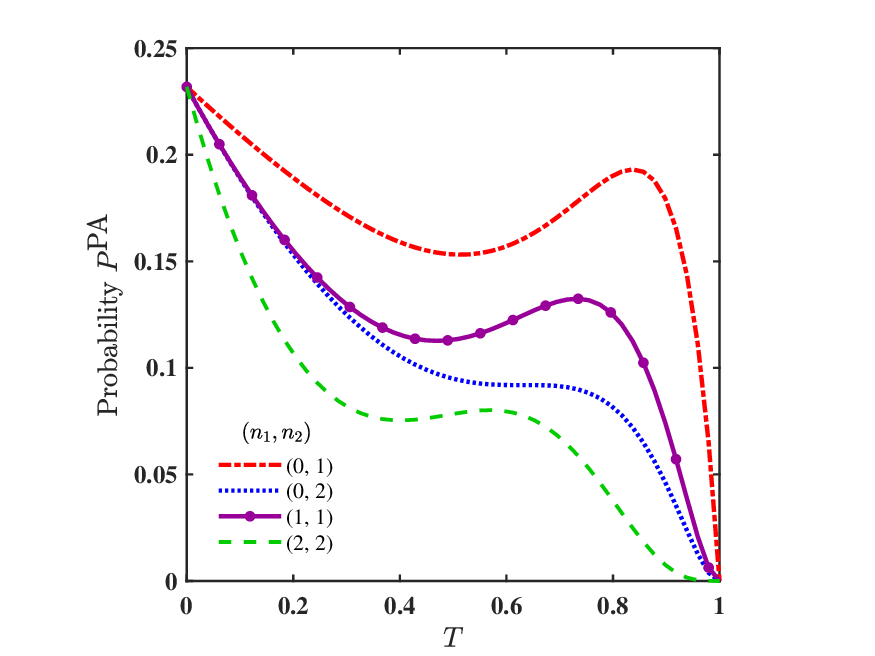}
		\caption{Success probability $P^{\mathrm{PA}}$ as a function of the transmissivity $T$ ($T_1=T_2=T$) for different photon addition configurations at fixed squeezing parameter $\xi=0.8$ and with the initial Fock number $(m_1,m_2)=(1,1)$.}
		\label{Prob_T}
	\end{figure}
	The success probability of the photon addition operation exhibits a strong dependence on both the squeezing parameter and the beam-splitter transmissivity as shown in Figs.~\ref{Prob_r} and \ref{Prob_T}. In Fig.~\ref{Prob_r}, for a fixed transmissivity $T_1=T_2=0.9$, 
	the success probability initially increases with the squeezing parameter $\xi$, attains its maximum, and subsequently decreases. The asymmetric single-photon addition configuration $(0,1)$ corresponds to the highest success probability, whereas higher-order photon addition configurations yield lower probabilities. In Fig.~\ref{Prob_T}, the success probability shows a nonmonotonic dependence on the transmissivity $T$. For the different addition patterns, the probability initially decreases, followed by a rise over an intermediate range of $T$, before vanishing as $T\rightarrow1$. This decline reflects the increasingly low probability of registering the required ``off'' detection events in the high-transmissivity regime. The lower-order photon addition again demonstrates higher success probabilities  than higher-order operations. These results affirm that the success probability of photon addition depends on the interplay among the squeezing parameter, beam-splitter transmissivity, and the number of added photons, emphasizing the importance of appropriate parameter optimization for efficient PA-TMSFS generation.
	
	Beyond its role in determining the success probability, the Wigner characteristic function given in Eq. \eqref{PAWCF} provides a unified description of multiple relevant quantum states as special limiting cases. In particular, substituting $m_1=m_2=0$ recovers the WCF of the PA-TMSV state, while setting both photon addition and input Fock-state indices to zero yields the TMSV state. These limiting cases further establish the general applicability of the present formulation.
	\section{Fidelity Analysis of PA-TMSFS}
	\label{fidelity}
	
\begin{figure}[htbp]
		\centering
		\includegraphics[trim=70pt 520pt 50pt 74pt,clip,width=0.5\textwidth]{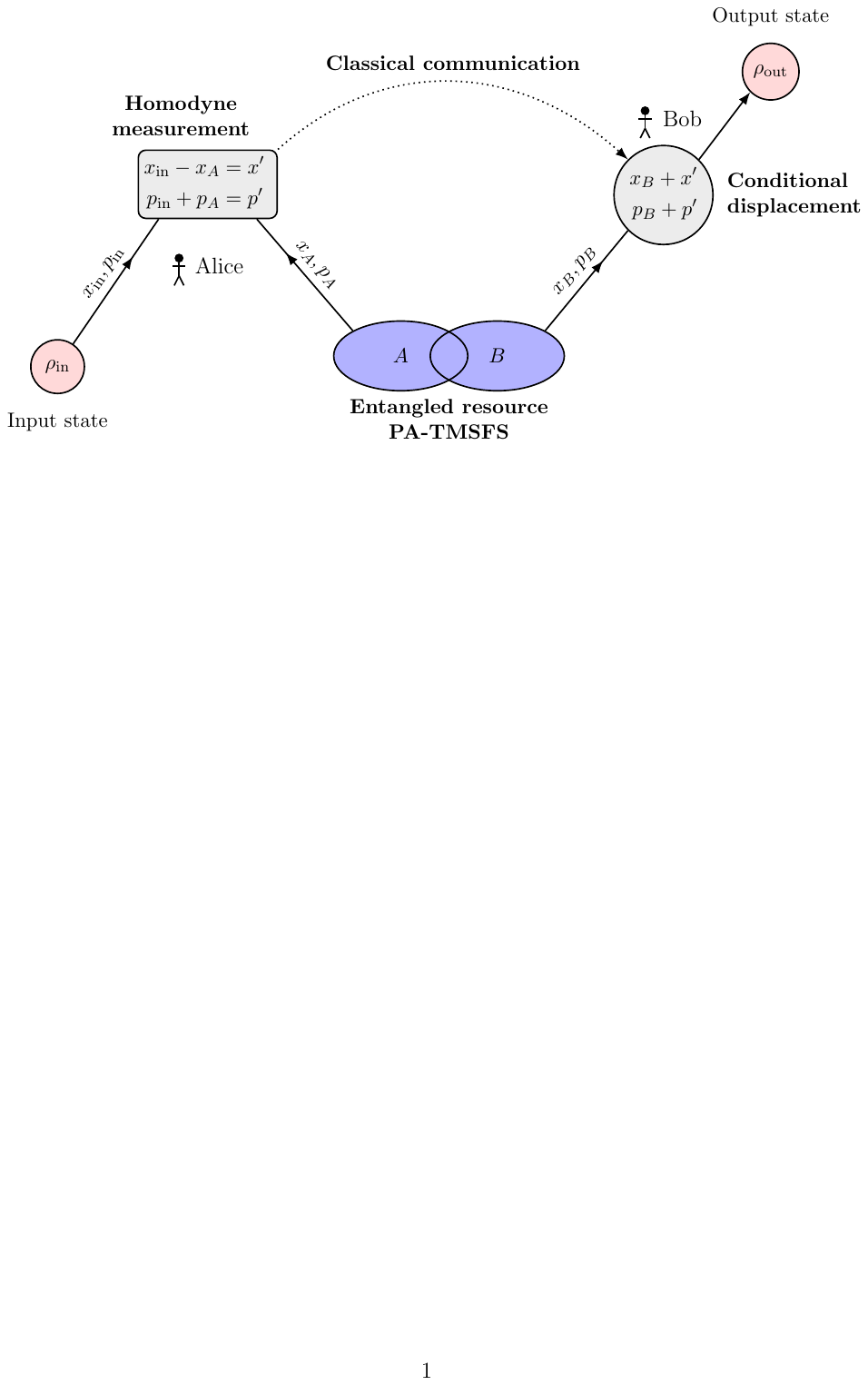}		
		\caption{Schematic diagram of CV quantum teleportation.}
		\label{QT_scheme}
	\end{figure}

	To assess the usefulness of PA-TMSFS state as a quantum resource, we consider its application to CV quantum teleportation within the Braunstein-Kimble (BK) protocol \cite{Braunstein1998}. A schematic representation of the teleportation protocol is shown in Fig.~\ref{QT_scheme}.
	
	The two modes of the PA-TMSFS resource, denoted by $A$ and $B$, are shared between Alice and Bob. The resource is characterized by the density operator $\rho_{A'_1A'_2}$ and its corresponding Wigner characteristic function $\chi_{A'_1A'_2}^{\mathrm{PA}}(\Lambda)$. Alice combines the unknown input state $\rho_{\mathrm{in}}$, 
	described by the characteristic function $\chi_{\mathrm{in}}(\Lambda_{\mathrm{in}})$, with mode $A$ and performs conjugate-quadrature homodyne measurements. The outcomes $(x',p')$ are then communicated to Bob, who applies a conditional displacement to mode $B~(x_B,p_B)$ to obtain the teleported state $\rho_{\mathrm{out}}$. Within the phase-space description of the BK protocol, the characteristic function of this output state can be expressed in terms of the characteristic functions of the input state and the shared entangled resource \cite{Marian2006}. In particular, it is given by
	\begin{equation}
		\chi_{\mathrm{out}}(\tau_2,\sigma_2)
		=
		\chi_{\mathrm{in}}(\tau_2,\sigma_2)
		\chi_{\mathrm{res}}(\tau_2,-\sigma_2,\tau_2,\sigma_2),
	\end{equation}
	where $\chi_{\mathrm{res}}$ represents the Wigner characteristic function of the PA-TMSFS resource. The teleportation fidelity, which quantifies the overlap between the input and teleported states, is defined as
	\begin{equation}
		F=\mathrm{Tr}\left[\rho_{\mathrm{in}}\rho_{\mathrm{out}}\right],
	\end{equation}
	with the characteristic function formulation as \cite{Chizhov2002}
	\begin{equation}
		F=
		\frac{1}{2\pi}
		\int d^2\Lambda\,
		\chi_{\mathrm{in}}(\Lambda)\,
		\chi_{\mathrm{out}}(-\Lambda).
		\label{F_B}
	\end{equation}	
	The teleportation fidelity characterizes the quality of the reconstructed state with $F=1$ corresponds to perfect teleportation. For coherent state inputs, $F=1/2$ represents the classical limit \cite{Braunstein2000TF,Braunstein2001} whereas $F=2/3$ defines the no-cloning threshold \cite{Cerf2000,Grosshans2001}. We now employ Eq.~\eqref{F_B} to investigate the teleportation fidelity associated with Gaussian and non-Gaussian input states, using these benchmarks as reference points. This analysis allows us to elucidate how photon addition influences teleportation fidelity across both classes of input states.
	\subsection{Gaussian state teleportation}	
	We first investigate the teleportation performance of the engineered TMSFS resource, namely PA-TMSFS, for Gaussian input states. Coherent and squeezed vacuum states are considered as representative Gaussian inputs, both widely employed for characterizing CV teleportation performance. We next obtain the fidelity expressions for teleporting coherent and squeezed states and examine their dependence on the resource squeezing parameter and channel transmissivity.
	\subsubsection{Coherent state}
	We consider the teleportation of a single-mode coherent state using PA-TMSFS as the entangled resource. Substituting the Wigner characteristic functions of the coherent state [Eq.~\eqref{WCF_CS}] and the PA-TMSFS resource [Eq.~\eqref{PAWCF}] into Eq.~\eqref{F_B}, the teleportation fidelity is obtained as
	\begin{align}
		F = \frac{\hat{F}}{d_0}\exp\left(u^TM_4u\right),
		\label{FCS}
	\end{align}
	where $\hat{F}$ denotes the differential operator and the matrix $M_4$ is given in Eq.~\eqref{M_4} of Appendix~\ref{App}. This expression accounts for the combined effects of the resource squeezing, beam-splitter transmissivities, and photon addition operations, thereby providing a systematic investigation of their impact on teleportation fidelity.
	
	We further analyze the fidelity as a function of the squeezing parameter and transmissivity for different photon addition configurations, considering the parent TMSFS as a reference. This comparison provides a measure of the improvement in teleportation performance achieved through photon addition over the underlying TMSFS.
	\begin{figure}[htbp]
		\centering
		\includegraphics[trim=0pt 0pt 0pt 0pt,clip,width=0.5\textwidth]{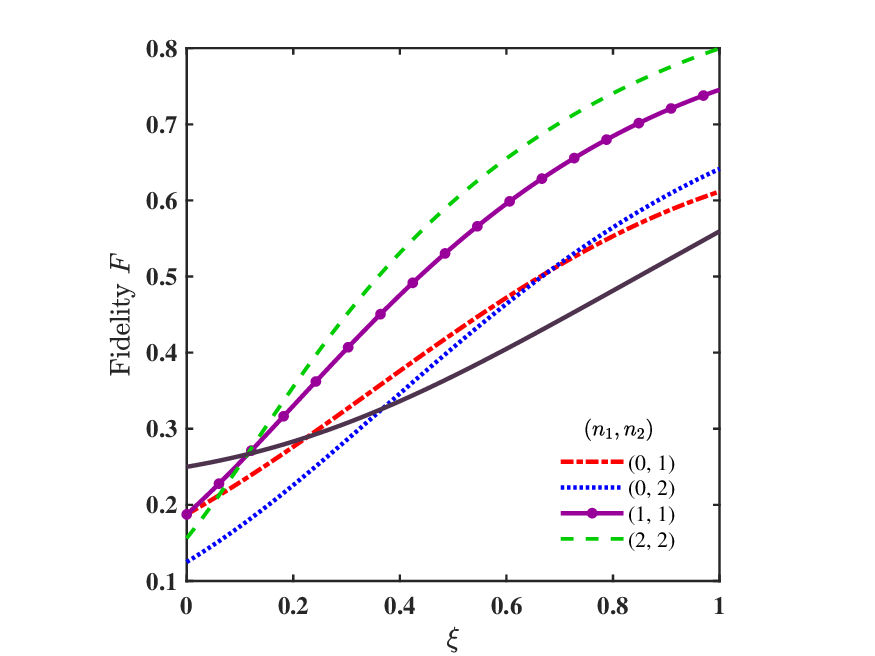}
		\caption{Teleportation fidelity $F$ for the input coherent state as a function of the squeezing parameter $\xi$ for different photon addition configurations $(n_1, n_2)$. The underlying Fock state is fixed at $(m_1,m_2)=(1,1)$ with $T_1=T_2=0.9$. The solid black line represents the fidelity obtained for teleporting coherent state using the TMSFS resource.}
		\label{FCoherent_r}
	\end{figure}
	\begin{figure}[htbp]
		\centering
		\includegraphics[trim=0pt 0pt 0pt 0pt,clip,width=0.5\textwidth]{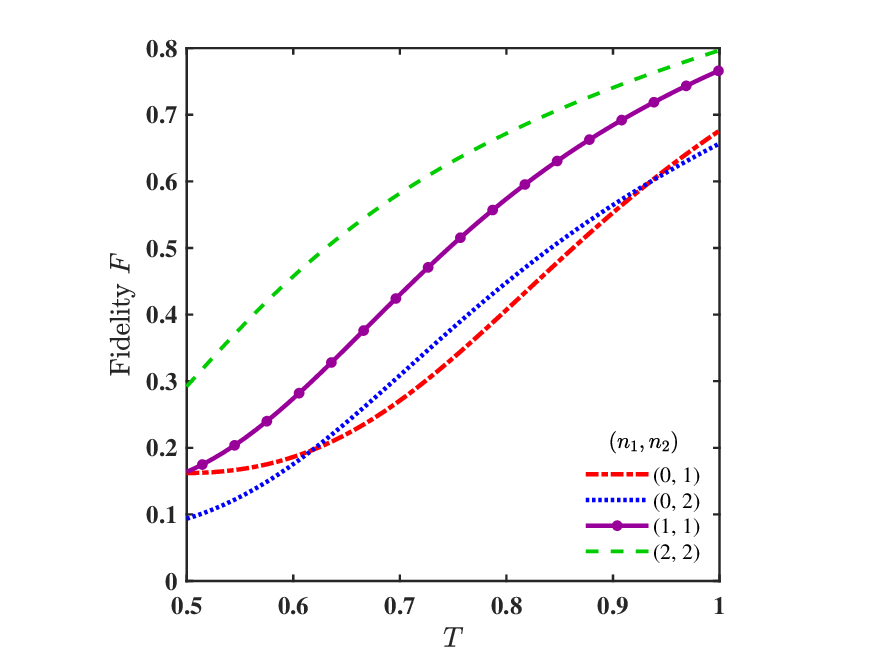}
		\caption{Teleportation fidelity $F$ as a function of the beam-splitter transmissivity $T$ ($T_1=T_2=T$) for different photon addition configurations of the PA-TMSFS channel. The underlying Fock state is fixed at $(m_1,m_2)=(1,1)$ with $\xi=0.8$.}
		\label{FCoherent_T}
	\end{figure}
	Figs.~\ref{FCoherent_r} and \ref{FCoherent_T} show the fidelity for coherent state teleportation as a function of the squeezing parameter $\xi$ and the beam-splitter transmissivity $T$, respectively, for different photon addition configurations $(n_1, n_2)$ of the PA-TMSFS resource, benchmarked against the TMSFS channel. In Fig.~\ref{FCoherent_r}, the fidelity increases monotonically with the squeezing parameter for all configurations, with the symmetric cases $(1,1)$ and $(2,2)$ exhibiting the steepest increase and outperforming the parent TMSFS channel over most of the parameter range. Among all the cases, $(2,2)$ yields the highest fidelity throughout, approaching a maximum value of $F\approx0.8$ at large $\xi$. The asymmetric configurations $(0,1)$ and $(0,2)$ initially exhibit lower fidelities but surpass the TMSFS resource at intermediate squeezing. A similar behavior is observed with increasing $T$ as shown in Fig.~\ref{FCoherent_T}. The $(2,2)$ configuration again provides the highest fidelity among the different photon added resources, followed by the $(1,1)$ pattern, while the asymmetric configurations remain comparatively less effective. For the entangled TMSFS channel, the teleportation fidelity is independent of $T$ and remains constant at $F=0.48$, as no beam-splitter-based photon addition operation is involved. As $T$ increases, the photon added configurations gradually surpass the reference TMSFS state. The symmetric configurations exhibit an earlier crossover and maintain the highest fidelity advantage as $T$ approaches unity, while the asymmetric configurations exceed the reference only at relatively higher transmissivities. 
	
	These results show that symmetric photon addition, particularly the $(2,2)$ configuration, offers a consistent advantage in coherent state teleportation fidelity over the Gaussian TMSFS resource across both parameter regimes. In contrast, asymmetric patterns $(0,1)$ and $(0,2)$ surpass the TMSFS benchmark only at sufficiently high values of $\xi$ or $T$.
	\subsubsection{Squeezed vacuum state}
	We next turn to the teleportation of a single-mode squeezed vacuum, a nonclassical input state in which quantum noise is distributed unevenly between the two quadratures. This asymmetry makes the squeezed state considerably more susceptible to imperfections in the entangled channel than a coherent state, so that the resulting fidelity provides a deeper insight into the potential of the PA-TMSFS resource for preserving the nonclassical features of the input squeezed state $\ket{\epsilon}=S(\epsilon)\ket{0}$.
	
	Substituting the WCF of the squeezed vacuum state [Eq.~\eqref{WCF_SS}] and that of the PA-TMSFS resource into Eq.~\eqref{F_B}, the fidelity reduces to
	\begin{align}
		F=	\frac{\hat{F}}{e'_0}
		\exp\left(u^{T} M_5
		u\right),
		\label{FSS}
	\end{align}
	with $e'_0= \sqrt{e_0}\,(\beta-\alpha t_1 t_2)$, $\hat{F}$ is the associated differential operator, and the matrix $M_5$ is defined in Eq.~\eqref{M_5}. The fidelity is then studied as a function of the squeezing $\xi$ and the beam-splitter transmittivity $T$.
	\begin{figure}[htbp]
		\centering
		\includegraphics[trim=0pt 0pt 0pt 0pt,clip,width=0.5\textwidth]{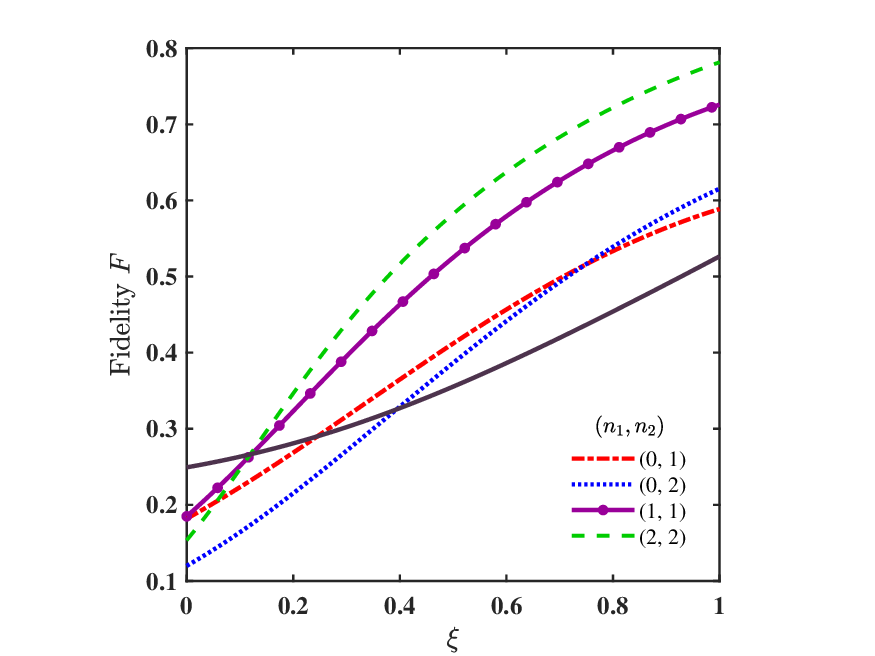}
		\caption{Teleportation fidelity $F$ as a function of the squeezing parameter $\xi$ for teleporting  squeezed vacuum state $\ket{\epsilon}$ with $\epsilon=0.3$. The source Fock state is fixed at $(m_1,m_2)=(1,1)$ with $T_1=T_2=0.9$. The solid black line represents the fidelity obtained using the TMSFS resource for teleporting squeezed state.}
		\label{FSqueezed_r}
	\end{figure}
	\begin{figure}[htbp]
		\centering
		\includegraphics[trim=0pt 0pt 0pt 0pt,clip,width=0.5\textwidth]{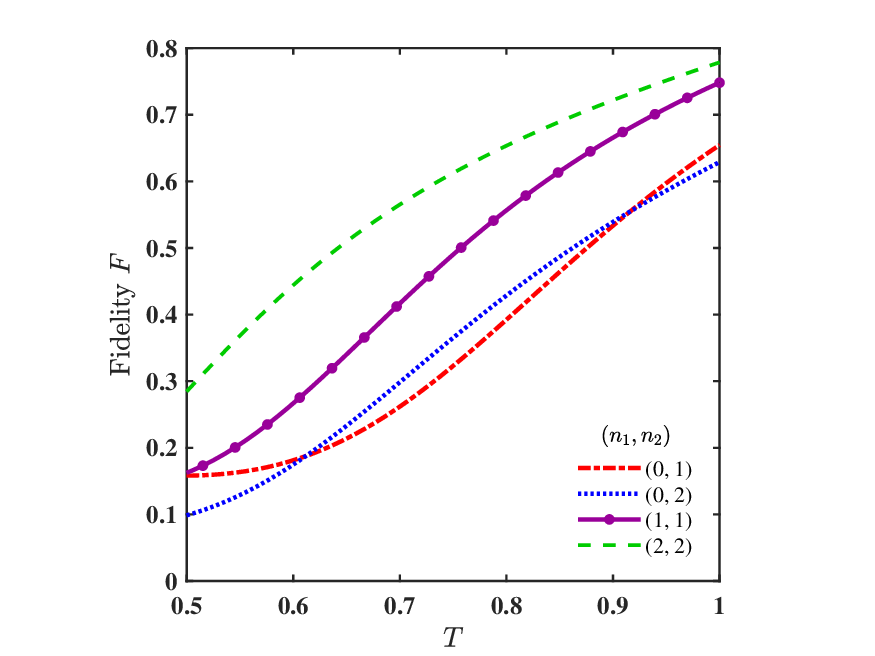}
		\caption{Teleportation fidelity $F$ as a function of the beam-splitter transmissivity $T$ for an input squeezed vacuum state with different photon addition configurations. The resource state is fixed at $(m_1,m_2)=(1,1)$ with $\xi=0.8$, while the input squeezing parameter is $\epsilon=0.3$. 
		}
		\label{FSqueezed_T}
	\end{figure}
	
	Figs.~\ref{FSqueezed_r} and \ref{FSqueezed_T} show the teleportation fidelity as a function of the squeezing parameter $\xi$ and the beam-splitter transmissivity $T$, respectively, for different photon addition configurations. For increasing $\xi$, the fidelity for all photon added configurations increases, with the symmetric $(2,2)$ and $(1,1)$ configurations providing the most significant enhancement at higher squeezing. In contrast, the asymmetric configurations $(0,1)$ and $(0,2)$ exhibit comparatively lower fidelities. Although the TMSFS resource provides higher fidelity in the low-squeezing regime, the $(2,2)$ and $(1,1)$ configurations surpass it as the squeezing increases, demonstrating the advantage of photon addition in the stronger squeezing regime. In Fig.~\ref{FSqueezed_T}, the TMSFS resource is independent of $T$ and thus corresponds to a fixed fidelity value of $F=0.45$. As $T$ increases, all four photon addition configurations eventually surpass this constant reference fidelity, with the symmetric configurations crossing it earlier and achieving the overall highest fidelity as $T\to1$. The asymmetric configurations outperform the parent TMSFS only at comparatively higher transmissivity and remain below the symmetric curves throughout.
	
	Overall, these results are consistent with those obtained in the coherent state case. While the underlying TMSFS resource can perform better at low squeezing or low transmissivity, symmetric photon addition, particularly the $(2,2)$ configuration, provides a clear advantage in the higher-parameter regime. These findings establish that symmetric photon addition offers more effective and robust strategy for teleporting the Gaussian states. We next explore whether the symmetric photon addition retains its advantage while teleporting non-Gaussian states.
	\subsection{Non-Gaussian state teleportation}
	Following the investigation of Gaussian state teleportation, we extend our analysis to non-Gaussian inputs, namely Fock and Schr\"odinger cat states. Unlike Gaussian inputs, these states may exhibit negative regions in their Wigner functions and are characterized by higher-order phase-space moments, making them more sensitive to the entanglement and non-Gaussianity features of the PA-TMSFS resource. Investigating their teleportation performance therefore allows us to determine whether the benefits of photon addition persist when the input state exhibits non-Gaussian characteristics. As non-Gaussian inputs are more sensitive to channel losses, a high transmissivity ($T=0.99$) is used throughout the analysis to approximate near ideal photon addition and isolate the resource's intrinsic performance. In the following subsection, we investigate the fidelity for an input Fock state, and subsequently for an input Schr\"odinger cat state using the same characteristic function approach employed earlier.
	\subsubsection{Fock state}
	We next consider the teleportation of a Fock state $\ket{n}$ employing the PA-TMSFS channel. Using the corresponding Wigner characteristic function [Eq.~\eqref{Lag}] along with that of the PA-TMSFS resource, the fidelity takes the form
	\begin{align}
		F &=	\frac{\hat{F}}{a_0}
		\sum_{k=0}^{n}
		\sum_{l=0}^{n}
		\frac{
			\binom{n}{k}
			\binom{n}{l}
		}
		{k!\,l!}
		\frac{\partial^{k+l}}
		{\partial z^{k+l}}
		\left[
		\frac{1}{z}
		\exp\!\left(
		u^{T} M_4 u
		\right)
		\right],
		\label{FFS}
	\end{align}
	where $\hat{F}$ is the associated differential operator, $n$ is the photon number in the Fock state input. The fidelity continues to depend on the same set of resource parameters, namely the squeezing $\xi$, the transmissivities $T_1,T_2$, and the photon-addition indices $(n_1, n_2)$. We now examine how these parameters contribute to the fidelity for different addition configurations.
	\begin{figure}[htbp]
		\centering
		\includegraphics[trim=0pt 0pt 0pt 0pt,clip,width=0.5\textwidth]{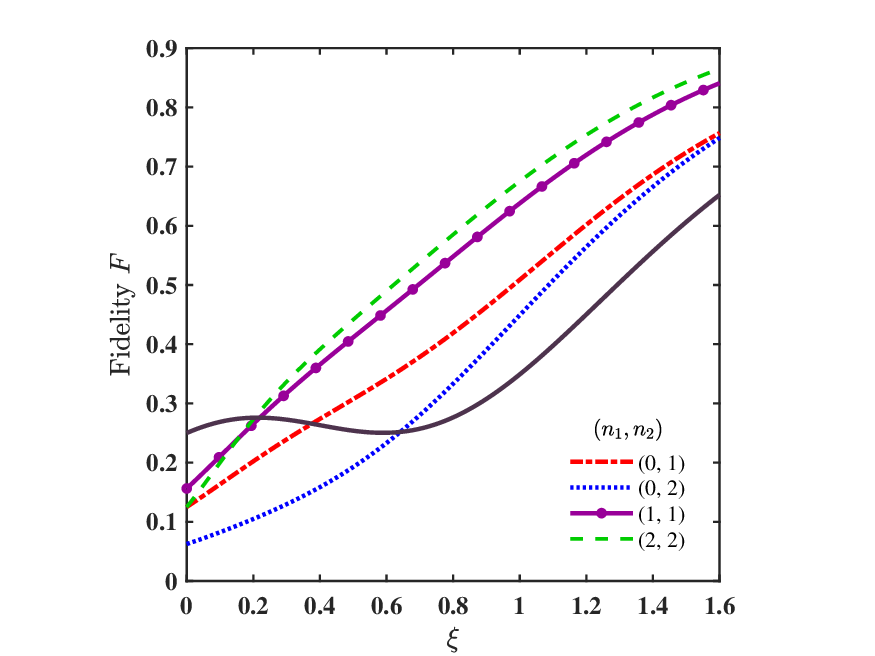}
		\caption{Teleportation fidelity $F$ for an input Fock state $\ket{1}$ as a function of the squeezing parameter $\xi$ for different photon addition configurations. The resource state is fixed at $(m_1,m_2)=(1,1)$ with $T_1=T_2=0.99$. The solid black line represents the fidelity obtained for teleporting Fock state $\ket{1}$ using the TMSFS channel.}
		\label{FFock_r}
	\end{figure}
	
	\begin{figure}[htbp]
		\centering
		\includegraphics[trim=0pt 0pt 0pt 0pt,clip,width=0.5\textwidth]{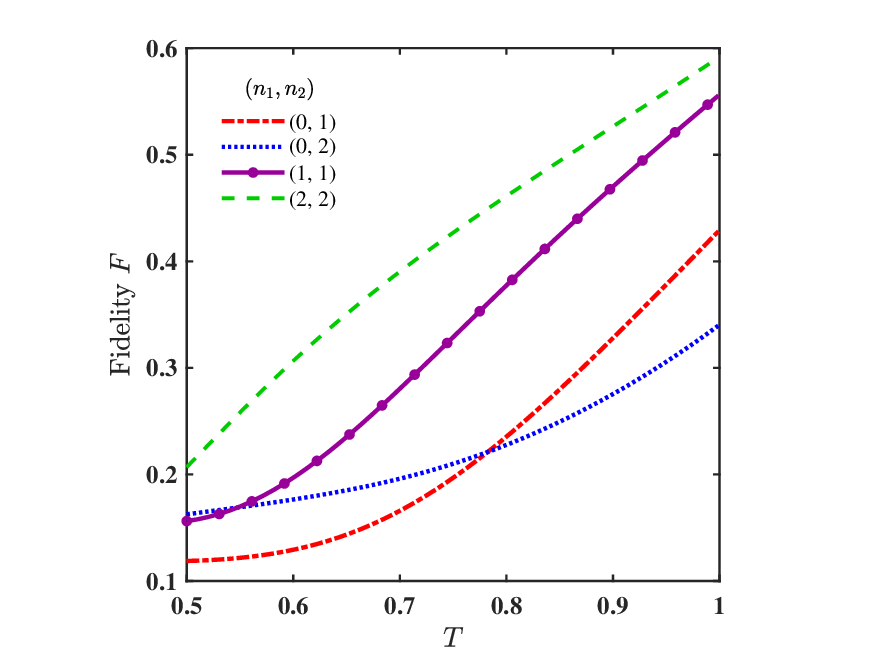}
		\caption{Teleportation fidelity $F$ for an input single-photon Fock state $\ket{1}$ as a function of the beam-splitter transmissivity $T$ ($T_1=T_2=T$) for different photon addition configurations. The resource state is fixed at $(m_1,m_2)=(1,1)$ with $\xi=0.8$. }
		\label{FFock_T}
	\end{figure}
	Figs.~\ref{FFock_r} and \ref{FFock_T} present the teleportation fidelity for the Fock state input $\ket{1}$ as a function of $\xi$ and $T$, respectively. The results highlight a clear distinction between symmetric and asymmetric photon addition schemes, while exhibiting an overall trend consistent with that observed for Gaussian inputs, with symmetric photon addition configurations outperforming the asymmetric ones. Here the $(2,2)$ scheme provides the highest fidelity ($F\approx0.86$) at $\xi=1.6$.
	A notable difference from the Gaussian input cases is observed for the unengineered TMSFS resource. The fidelity displays a non-mnonotonic variation with the squeezing parameter $\xi$, it increases initially, followed by a decline around intermediate regime, before increasing again at stronger squeezing. The fidelity for TMSFS remains below the classical benchmark over a substantial portion of the considered squeezing range, exceeding this limit only at higher squeezing, suggesting that enhanced entanglement in the teleportation channel is required to achieve high-fidelity transmission of non-Gaussian states. A similar dependence on the photon addition configuration is observed with increasing $T$. The symmetric configurations $(2,2)$ and $(1,1)$ again yield the higher fidelities, whereas the asymmetric configurations provide weaker enhancement and remain below the classical limit $F=1/2$ throughout $T$ range. The TMSFS fidelity is independent of $T$ and remains constant at $F=0.28$, and is therefore surpassed by the photon added configurations at moderate transmissivity. In contrast to the Gaussian input states, no configuration exhibits fidelity beyond the no-cloning limit across the chosen parameter range.
	
	In general, symmetric photon addition configurations demonstrate better teleportation performance, with the \((2,2)\) scheme emerging as the most effective among the configurations studied.  
	\subsubsection{Schr\"odinger cat state}
	We now evaluate the transmission of a Schr\"odinger cat state, described by the superposition of coherent states $\ket{\pm\gamma}$. The teleportation fidelity is obtained by substituting the Wigner characteristic functions of the cat state [Eq.~\eqref{WCF_CAT}] and the PA-TMSFS into Eq.~\eqref{F_B} as
	\begin{align}
		F
		&=
		\frac{\hat{F}}
		{d_0
			\left(
			1\pm e^{-2|\gamma|^{2}}
			\right)^{2}}
		\Biggl\{
		\exp\left(
		u^{T}M_4u
		\right)
		\nonumber\\
		&\quad\times
		\Bigg(
		\frac{1+e^{-4|\gamma|^{2}}}{2}
		+
		\frac{1}{2}
		\exp\left[
		-\frac{4a_0|\gamma|^{2}}{d_0}
		\right]
		\cosh\left[
		\frac{2(ipR+qS)}{d_0}
		\right]
		\nonumber\\
		&\qquad+
		\frac{e^{-4|\gamma|^{2}}}{2}
		\exp\left[
		\frac{4a_0|\gamma|^{2}}{d_0}
		\right]
		\cosh\left[
		\frac{2(qR+ipS)}{d_0}
		\right]
		\nonumber\\
		&\qquad\pm
		e^{-2|\gamma|^{2}}
		\cosh\left[
		\frac{qR+ipS}{d_0}
		\right]
		\cosh\left[
		\frac{qS+ipR}{d_0}
		\right]
		\Bigg)
		\Biggr\},
		\label{FCAT}
	\end{align}
	where $\hat{F}$ denotes the differential operator and  $\gamma=(q+ip)/\sqrt{2}$ is the amplitude of the coherent state. The effects of the squeezing parameter and the beam-splitter transmissivities are further analyzed for different photon addition configurations.
	\begin{figure}[htbp]
		\centering
		\includegraphics[trim=0pt 0pt 0pt 0pt,clip,width=0.5\textwidth]{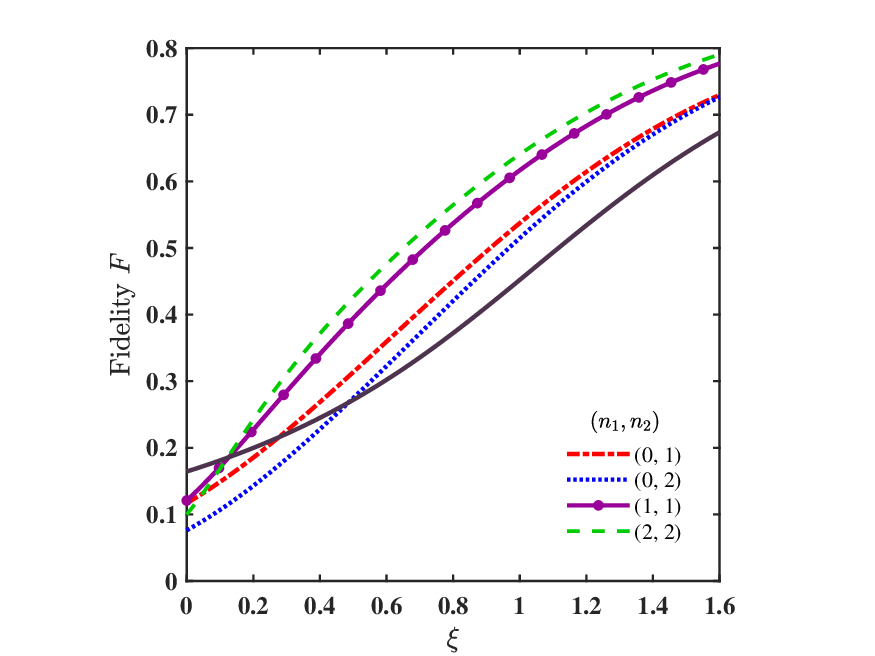}
		\caption{Teleportation fidelity $F$ for an input even Schr\"odinger cat state $\ket{\psi_{+}}$ as a function of the squeezing parameter $\xi$. The resource state is fixed at $(m_1,m_2)=(1,1)$ with $\gamma=1$ and $T_1=T_2=0.99$. The solid black line denotes the teleportation fidelity for cat state input using TMSFS resource.}
		\label{FCat_r}
	\end{figure}
	\begin{figure}[htbp]
		\centering
		\includegraphics[trim=0pt 0pt 0pt 0pt,clip,width=0.5\textwidth]{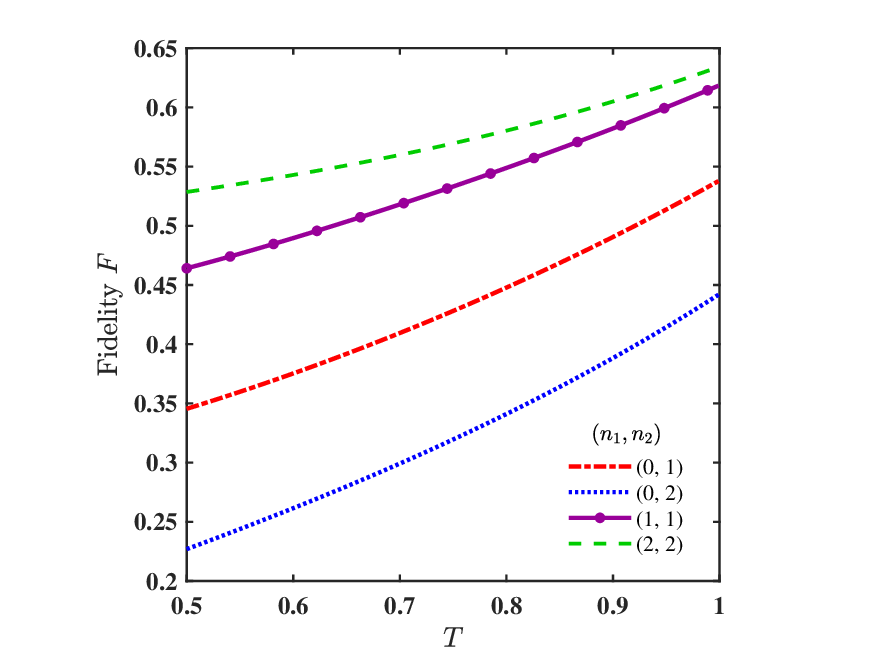}
		\caption{Teleportation fidelity $F$ for an input even Schr\"odinger cat state as a function of the beam-splitter transmissivity $T$ ($T_1=T_2=T$) for different photon addition configurations. The resource state is fixed at $(m_1,m_2)=(1,1)$ with $\gamma=1$ and $\xi=0.8$.}
		\label{FCat_T}
	\end{figure}
	
	Figs.~\ref{FCat_r} and \ref{FCat_T} show the teleportation fidelity for the input even Schr\"odinger cat state. As $\xi$ increases, all photon added configurations exhibit an increase in fidelity, with the symmetric configurations $(1,1)$ and $(2,2)$ providing the better performance. The $(2,2)$ scheme attains the highest fidelity over most of the considered range, closely followed by $(1,1)$. The asymmetric configurations $(0,1)$ and $(0,2)$ show comparatively lower fidelities, while the TMSFS reference initially provides a higher fidelity but is eventually surpassed by the photon added resources as squeezing increases. The dependence on $T$ shows a similar but more pronounced distinction between the configurations. The symmetric schemes again dominate, whereas $(0,1)$ and $(0,2)$ remain at comparatively lower fidelity values. Since TMSFS fidelity is independent of $T$ and remains fixed at $F=0.37$, photon addition offers a more substantial benefit as transmissivity increases. The results further confirm the superiority of symmetric photon addition in enhancing teleportation performance, with the \((2,2)\) configuration achieving the highest overall fidelity.
	
	As a whole, these results illustrate that the performance of PA-TMSFS resource is conditioned by the squeezing parameter, beam-splitter transmissivity, photon addition configuration, and the nature of the input state itself. For both Gaussian and non-Gaussian inputs, symmetric photon addition schemes consistently yield greater fidelity improvements relative to the unmodified TMSFS resource than their asymmetric counterparts. For Gaussian inputs, this enhancement is substantial within the squeezing range $\xi\in[0,1]$, with symmetric configurations providing the higher fidelities and $(2,2)$ scheme emerging as the best-performing configuration overall. Although, photon addition leads to improved fidelity for non-Gaussian inputs, the enhancement remains insufficient to exceed the no-cloning limit over the same squeezing range, even at high beam-splitter transmissivity; stronger squeezing is therefore required to achieve higher fidelity in this case.
	
	For a more comprehensive assessment of PA-TMSFS, we examine its teleportation fidelity relative to the parent TMSFS state and compare its performance with that of the PS-TMSFS resource studied previously \cite{Ankita2026PLA}. As the fidelity of the unengineered TMSFS also varies with the squeezing parameter, the absolute fidelity alone does not adequately capture the improvement obtained from the non-Gaussian operations. To facilitate this comparison, we define the fidelity enhancement as
	\begin{equation}
		\Delta F = F_{\mathrm{NG}}-F_{\mathrm{TMSFS}},
	\end{equation}
	where $F_{\mathrm{NG}}$ is the teleportation fidelity obtained using the non-Gaussian (photon-subtraced and photon-added) TMSFS resources. A positive value of $\Delta F$ indicates an improvement over the corresponding TMSFS resource, while  $\Delta F=0$ shows that non-Gaussian operation offers no advantage and $\Delta F<0$ indicates that the non-Gaussian resource performs less favorably than TMSFS.
	
	\section{Performance evaluation of Engineered TMSFS
}
\label{DelF}
To compare the effectiveness of photon addition and photon subtraction, we analyze the fidelity enhancement \(\Delta F\) and the probability-weighted fidelity enhancement \(P\Delta F\) for different classes of input states in this section. While \(\Delta F\) quantifies the variation in teleportation fidelity relative to the unengineered TMSFS resource, \(P\Delta F\) additionally accounts for the success probability of the corresponding non-Gaussian operation. 

To facilitate a direct comparison between photon subtraction and photon addition, the beam-splitter transmissivities are fixed at $T_1=T_2=0.9$ for all input states. In the high-transmissivity limit $T\rightarrow1$, the PS and PA curves corresponding to the same photon configuration become nearly indistinguishable for both classes of inputs. We consider the fidelity enhancement for the symmetric \((1,1)\) and \((2,2)\) patterns. The asymmetric configurations are not included in this comparison, given that their teleportation fidelities typically remain below the no-cloning threshold across the considered squeezing regime, limiting their practical advantage.

\subsection{Fidelity enhancement}	 
\label{F_Enhancement}

The fidelity enhancement $\Delta F$ is now compared for photon-subtracted and photon-added TMSFS resources for different classes of input states. We first examine coherent and squeezed vacuum inputs, followed by the non-Gaussian Fock and cat states. This comparison allows us to investigate how the nature of the input state influences the relative performance of the two non-Gaussian operations. 

\begin{figure}[htbp]
	\centering
	\begin{subfigure}{0.5\textwidth}
		\centering
		\caption{Coherent state input.}
		\includegraphics[trim=0pt 0pt 0pt 0pt,clip,width=\textwidth]{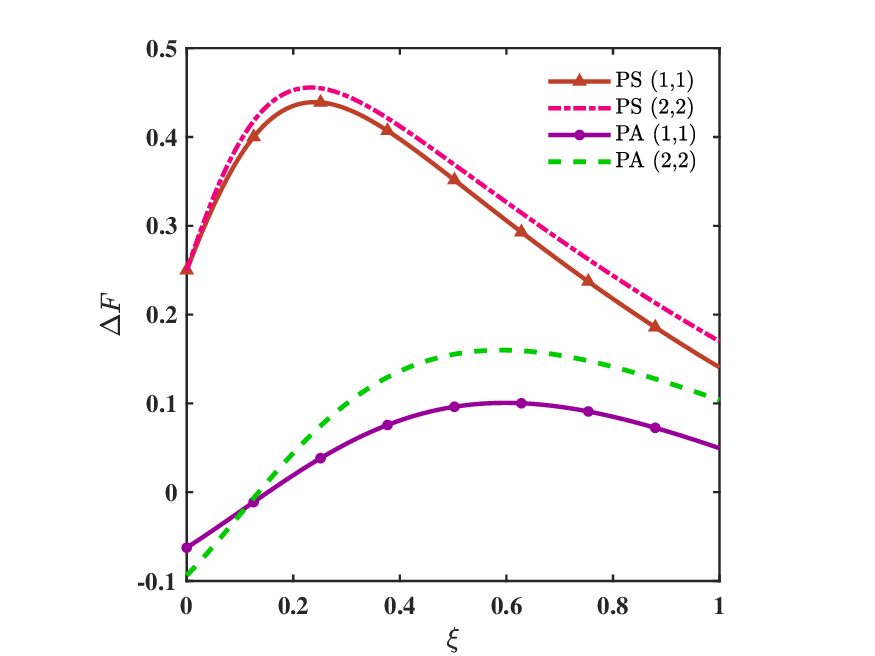}
		\label{CS_DF}
	\end{subfigure}
	\hfill
	\begin{subfigure}{0.5\textwidth}
		\centering
		\caption{Squeezed state input.}
		\includegraphics[trim=0pt 0pt 0pt 0pt,clip,width=\textwidth]{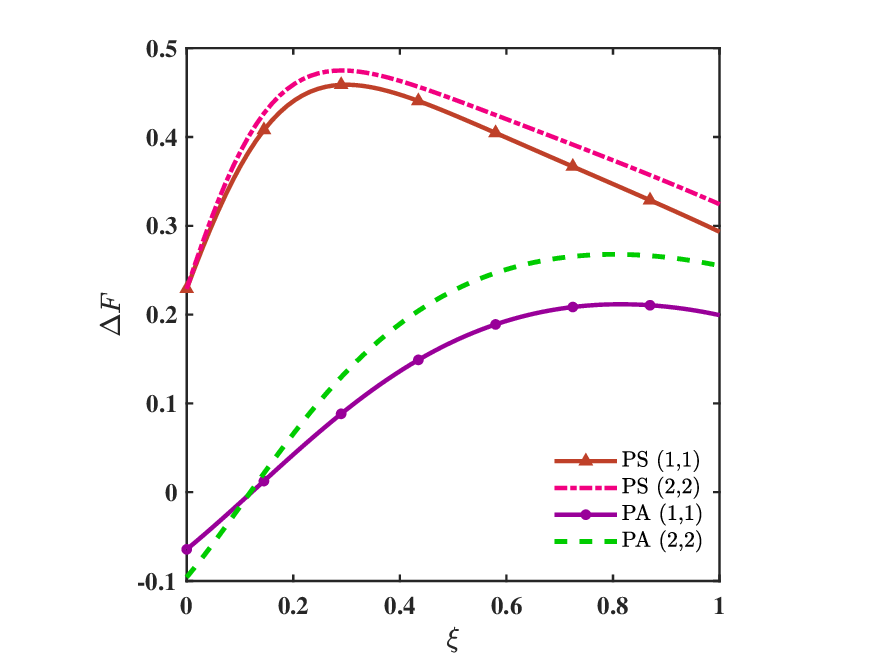}
		\label{SS_DF}
	\end{subfigure}
	
	\caption{Fidelity enhancement $\Delta F$ as a function of the squeezing parameter $\xi$ for symmetric photon subtraction and photon addition operations applied to the TMSFS resource. The input states are (a) coherent state and (b) squeezed state with $\epsilon=0.3$. The source Fock state is fixed at $(m_1,m_2)=(1,1)$ with $T_1=T_2=0.9$. }
	\label{fig:deltaF}
\end{figure}
Fig.~\ref{fig:deltaF} shows the fidelity enhancement $\Delta F$ for coherent and squeezed state inputs. For the coherent input [Fig.~\ref{CS_DF}], the two photon subtraction configurations $(1,1)$ and $(2,2)$ exhibit very similar behavior, with their fidelity enhancement attaining a maximum value around $\xi\simeq 0.2$. The corresponding photon addition configurations yield a considerably smaller enhancement, of which $(2,2)$ outperforms $(1,1)$ across most of the parameter range. A similar trend is observed for the squeezed state input [Fig.~\ref{SS_DF}]. The photon subtraction configurations again provide better advantage, while photon addition depicts a smaller but positive enhancement over a substantial part of the squeezing range. The difference between the two photon subtraction configurations remains marginal, whereas the $(2,2)$ photon addition pattern yields a substantial greater enhancement than the $(1,1)$ configuration. For both the Gaussian states, photon subtraction provides a larger fidelity enhancement than photon addition over the considered squeezing range. In each case, $\Delta F$ for the subtraction-based resource rises sharply, peaks at an intermediate value of $\xi$ before gradually decreasing at higher squeezing, indicating that photon subtraction is most advantageous in the intermediate squeezing regime.

\begin{figure}[htbp]
	\centering
	
	\begin{subfigure}{0.48\textwidth}
		\centering
		\caption{Fock state input.}
		\includegraphics[trim=0pt 0pt 0pt 0pt,clip,width=\textwidth]{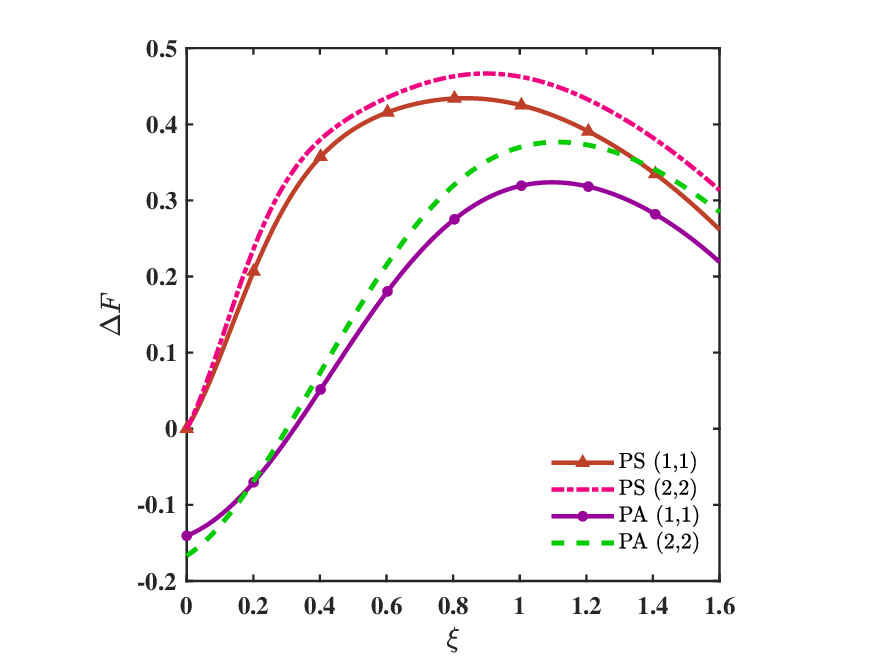}
	\end{subfigure}
	\hfill
	\begin{subfigure}{0.48\textwidth}
		\centering
		\caption{Cat state input.}
		\includegraphics[trim=0pt 0pt 0pt 0pt,clip,width=\textwidth]{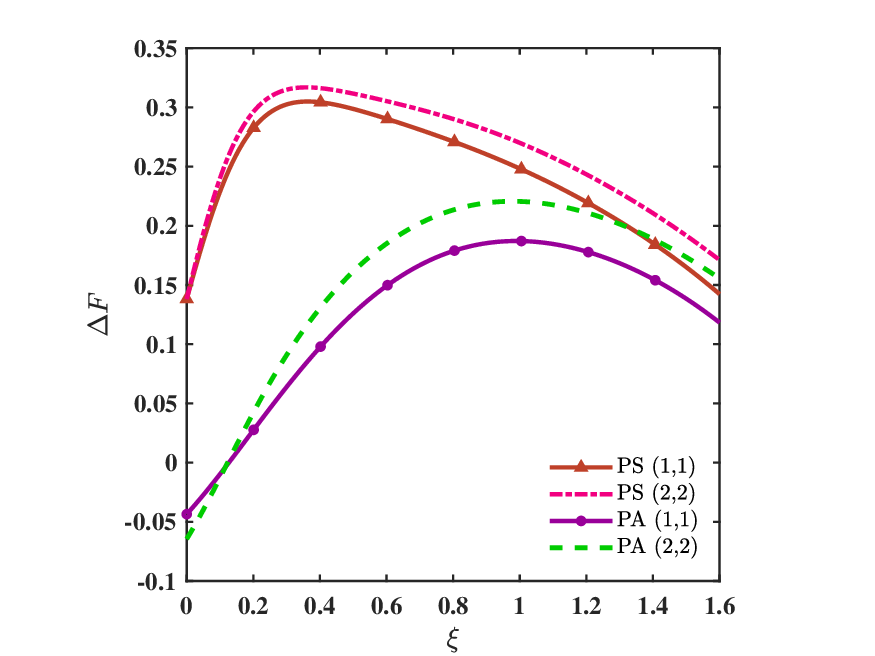}
	\end{subfigure}
	
	\caption{Fidelity enhancement $\Delta F$ as a function of the squeezing parameter $\xi$ for symmetric photon subtraction and photon addition operations applied to the TMSFS resource state for non-Gaussian input states: (a) Fock state $\ket{1}$ and (b) even Schr\"odinger cat state $\ket{\psi_{+}}$. The source Fock state is fixed at $(m_1,m_2)=(1,1)$ with $T_1=T_2=0.9$, while the cat state amplitude is fixed at $\gamma=1$.}
	\label{fig:deltaF_NG}
\end{figure}

A similar comparison is performed for the non-Gaussian Fock and cat state inputs as shown in Fig.~\ref{fig:deltaF_NG}.
For non-Gaussian input states, the fidelity enhancement follows a similar trend observed for Gaussian inputs: photon subtraction provides a greater enhancement in teleportation fidelity than photon addition over the considered range of the resource squeezing parameter. For both input states, the enhancement due to photon subtraction initially increases with $\xi$, reaches a maximum at an intermediate squeezing, and subsequently decreases as the squeezing is further increased. The photon addition configurations exhibit a same non-monotonic behavior, although the magnitude of the enhancement remains below than that obtained with photon subtraction. At sufficiently high squeezing, the enhancement for all photon configurations converges to nearly the same value, so that the relative advantage of photon subtraction over photon addition becomes negligible in this regime.


Altogether, these results indicate that the relative advantage of photon subtraction over photon addition persists irrespective of whether the input state is Gaussian or non-Gaussian. Moreover, the close behavior of the $(1,1)$ and $(2,2)$ PS configurations suggests that increasing the photon subtraction orders leads only to a marginal improvement in fidelity for the considered input states. 

Nevertheless, $\Delta F$ alone does not reflect the probabilistic character of the non-Gaussian operations. A substantial fidelity enhancement may come at the cost of a very low success probability, as reported for the PS-TMSV resource in the unit transmissivity limit~\cite{Chandan2023}. To account for this trade-off, we therefore consider the probability-weighted fidelity enhancement, defined as
\begin{equation}
	P \cdot\Delta F,
\end{equation}
where $P$ is the success probability of the corresponding non-Gaussian operation.
\subsection{Probability-weighted fidelity enhancement}	
\begin{figure}[htbp]
	\centering		
	\begin{subfigure}{0.48\textwidth}
		\centering
		\caption{Coherent state input}
		\includegraphics[trim=0pt 0pt 0pt 0pt,clip,width=\textwidth]{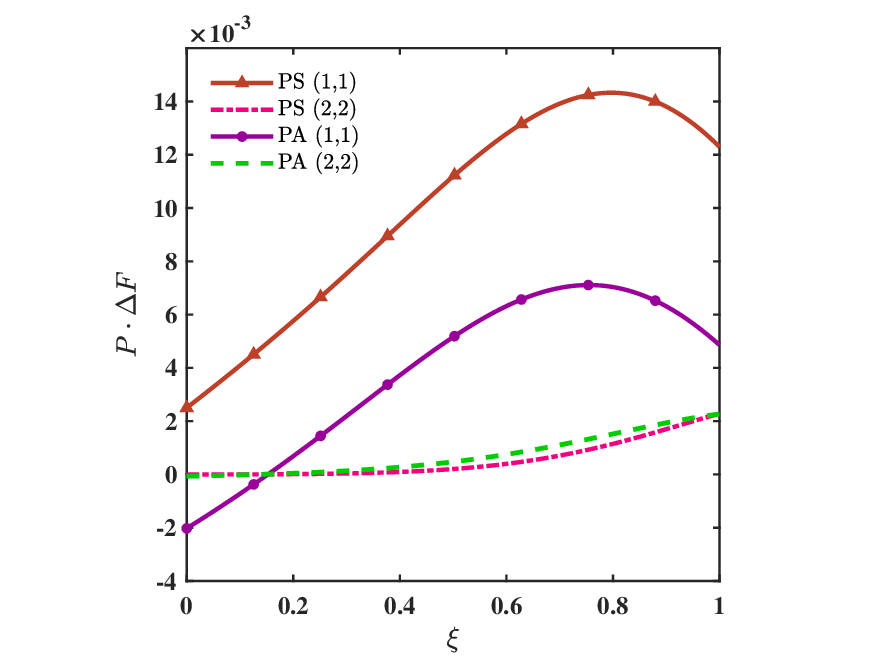}
		\label{CS_PDF}
	\end{subfigure}
	\hfill
	\begin{subfigure}{0.48\textwidth}
		\centering
		\caption{Squeezed state input}
		\includegraphics[trim=0pt 0pt 0pt 0pt,clip,width=\textwidth]{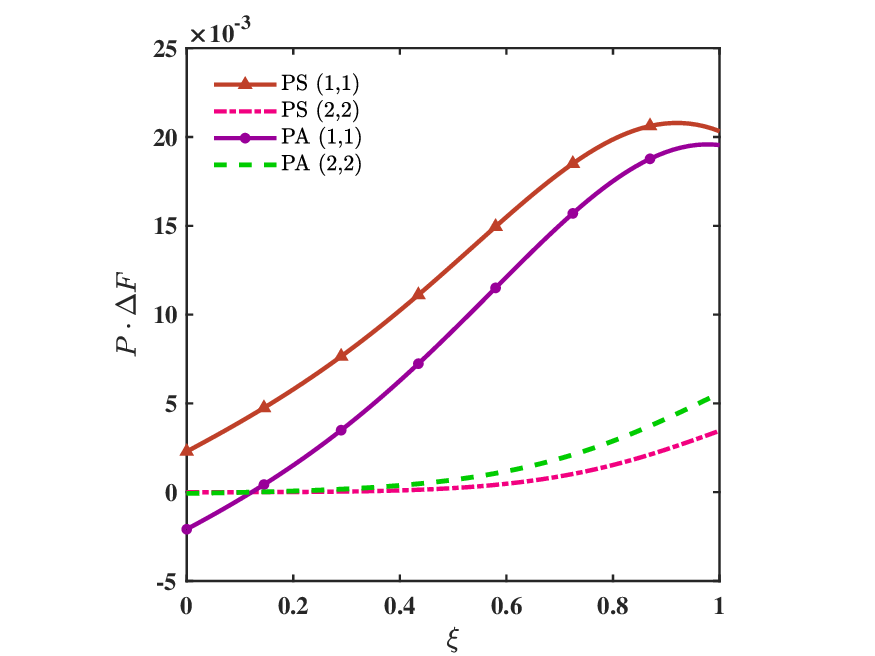}
		\label{SS_PDF}
	\end{subfigure}
	\caption{Probability-weighted fidelity enhancement $P\cdot \Delta F$ as a function of the squeezing parameter $\xi$ for symmetric PS- and PA-TMSFS for Gaussian input states: (a) coherent state and (b) squeezed state with $\epsilon=0.3$. The source Fock state is fixed at $(m_1,m_2)=(1,1)$ with $T_1=T_2=0.9$.}
	\label{fig:PdeltaF_G}
\end{figure}

To account for the probabilistic nature of resource preparation, we investigate the probability-weighted fidelity enhancement \(P\Delta F\) for coherent and squeezed vacuum inputs as a function of the squeezing parameter \(\xi\), as illustrated in Fig.~\ref{fig:PdeltaF_G}.

For both coherent and squeezed state inputs, photon subtraction and photon addition exhibit different behavior compared with that observed from $\Delta F$ alone. For the coherent state input [Fig.~\ref{CS_PDF}], the $(1,1)$ photon subtraction provides the largest value of $P \cdot \Delta F$ over the entire squeezing range considered. The corresponding gain increases with $\xi$, reaches a maximum at an intermediate squeezing of approximately $\xi\simeq0.8$, and decreases slightly thereafter. The $(1,1)$ photon addition configuration also yields a probability-weighted enhancement but remains substantially below that of photon subtraction. The $(2,2)$ configurations for both PS- and PA-TMSFS exhibit considerably lower values, due to the reduced success probability associated with higher-order photon addition and subtraction processes. For the squeezed state input [Fig.~\ref{SS_PDF}], the probability-weighted enhancement is conparatively larger than that obtained for the coherent state input. The $(1,1)$ photon subtraction configuration once again yields the maximum fidelity enhancement at low and intermediate squeezing, while the corresponding photon addition pattern gradually approaches its performance as $\xi$ increases. In addition, the $(2,2)$ configurations remain significantly below the corresponding $(1,1)$ cases throughout the considered range.

These results indicate that the relative merits of different configurations can change considerably when their success probabilities are taken into account alongside fidelity gains. While the fidelity enhancement plots show a clear advantage of photon subtraction over photon addition for the considered Gaussian inputs, the probability-weighted quantity manifests that the relative advantage is determined not only by the magnitude of $\Delta F$, but also by the probability with which the corresponding non-Gaussian state is successfully prepared. In particular, the symmetric $(1,1)$ photon subtraction configuration provides the most favorable probability-weighted enhancement for the coherent state input, whereas for the squeezed state input, both the PA- and PS- $(1,1)$ configurations approach one another at larger squeezing.

\begin{figure}[htbp]
	\centering		
	\begin{subfigure}{0.48\textwidth}
		\centering
		\caption{Fock state input}
		\includegraphics[trim=0pt 0pt 0pt 0pt,clip,width=\textwidth]{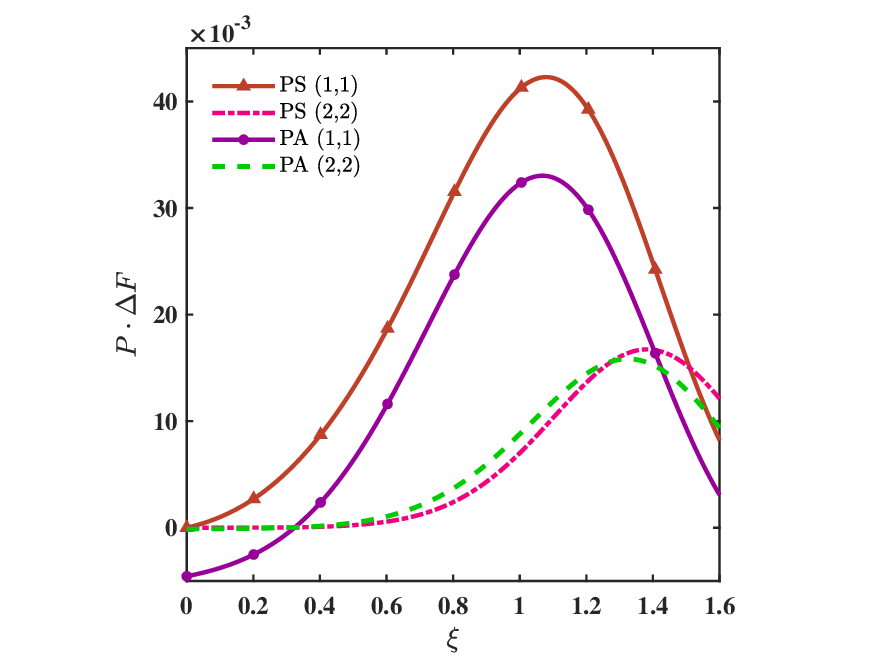}
		\label{FS_PDF}
	\end{subfigure}
	\hfill
	\begin{subfigure}{0.48\textwidth}
		\centering
		\caption{Cat state input}
		\includegraphics[trim=0pt 0pt 0pt 0pt,clip,width=\textwidth]{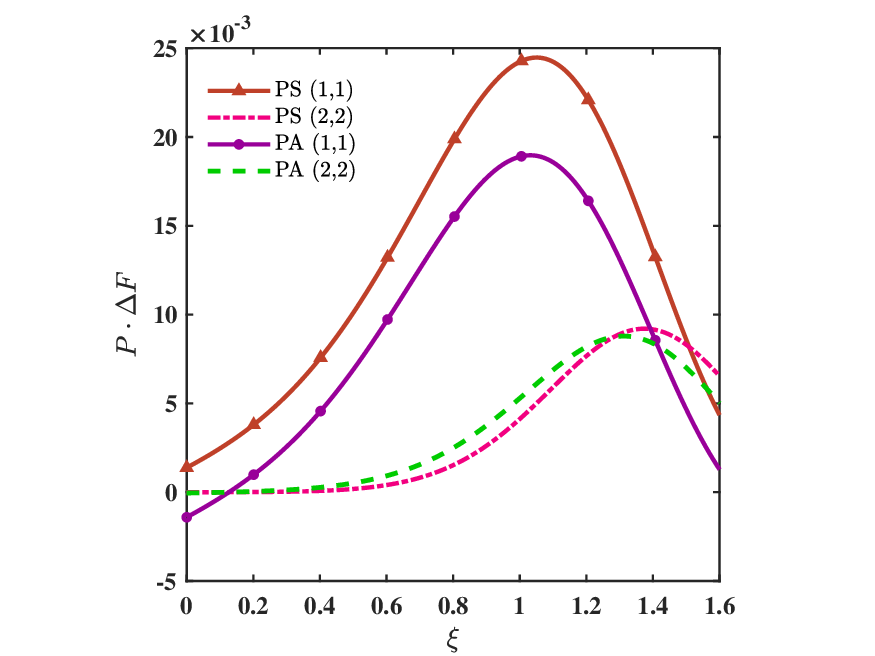}
		\label{CatS_PDF}
	\end{subfigure}
	\caption{Probability-weighted fidelity enhancement $P\cdot \Delta F$ as a function of the squeezing parameter $\xi$ for symmetric photon subtraction and photon addition operations applied to TMSFS state for non-Gaussian input states: (a) Fock state $\ket{1}$ and (b) even Schr\"odinger cat state $\ket{\psi_{+}}$. The source Fock state is fixed at $(m_1,m_2)=(1,1)$ with $T_1=T_2=0.9$, while the cat state amplitude is fixed at $\gamma=1$.}
	\label{fig:PdeltaF_NG}
\end{figure}

We next examine the probability-weighted fidelity enhancement for the non-Gaussian Fock and cat state inputs, as shown in Fig.\ref{fig:PdeltaF_NG}.

The probability-weighted fidelity enhancement for both non-Gaussian inputs exhibits a behavior similar to that observed for the Gaussian states. For the Fock state input [Fig.~\ref{FS_PDF}], the $(1,1)$ photon subtraction configuration yields the maximum advantage, followed by the $(1,1)$ photon addition scheme. Both of them exhibit a peak at intermediate squeezing before declining at higher values of $\xi$. The $(2,2)$ configurations provide considerably smaller enhancements, reflecting the reduced success probabilities associated with higher-order non-Gaussian operations.
A similar behavior is observed for the even Schrödinger cat state input [Fig.~\ref{CatS_PDF}]. The $(1,1)$ photon subtraction configuration again provides the largest probability-weighted enhancement, while photon addition yields a lower maximum. The $(2,2)$ configurations remain comparatively less effective. Thus for both non-Gaussian inputs, the $(1,1)$ photon subtraction configuration offers the most favorable balance between fidelity enhancement and success probability.

While photon subtraction provides a larger $P\cdot \Delta F$, the $(1,1)$ photon addition configuration can achieve comparable probability-weighted enhancement, particularly for non-Gaussian input states. However, this analysis considers only the conditional success probability of the non-Gaussian operation itself, neglecting the preparation of its ancillary states. Photon addition requires non-vacuum Fock states in the ancillary modes, whose preparation is itself probabilistic. Following the Fock state preparation scheme in \cite{Chandan2023}, the probability of generating a Fock state $\ket{n}$ is
\begin{equation}
	P_{\ket{n}}=(1-\lambda^2)\lambda^{2n},
\end{equation}
with $\lambda=\tanh r$. Consequently, the effective success probability of photon addition becomes
\begin{equation}
	P_{\mathrm{eff}}^{\mathrm{PA}}
	=P_{\ket{n_1}}P_{\ket{n_2}}P^{\mathrm{PA}}.
\end{equation}
For the symmetric $(1,1)$ photon addition configuration, this introduces an additional factor $P_{\ket{1}}^2$.  In contrast, photon subtraction requires vacuum ancillary inputs and therefore does
not incur this extra preparation penalty. Although, the PA-$(1,1)$ operation provides a compatible probability-weighted enhancement as compared with $(1,1)$ photon subtraction, this apparent advantage is further reduced when the probabilistic preparation of the ancillary states is taken into account. 
In aggregate, our results show that photon subtraction is the favorable non-Gaussian operation for enhancing CV quantum teleportation using the TMSFS resource. Considering fidelity enhancement, operational success probability, and ancillary state preparation requirements together, the symmetric $(1,1)$ photon subtraction offers the most beneficial trade-off among the operations considered in this work.

\section{Conclusion}	
In this work, we have studied the role of photon addition for engineering two-mode squeezed Fock states for CV quantum teleportation. We have obtained the WCF of PA-TMSFS, corresponding success probability, and evaluated the teleportation fidelity for coherent, squeezed vacuum, Fock, and even Schrödinger cat state inputs. By examining the dependence on the squeezing parameter, beam-splitter transmissivity, and photon addition configuration, we have characterized the performance of PA-TMSFS across different classes of input states.

We have further compared photon addition with photon subtraction by evaluating the fidelity enhancement relative to the parent TMSFS resource. For the symmetric configurations considered, particularly for $(
2,2)$, photon subtraction generally provides a larger fidelity enhancement than photon addition.
The results show that the relative performance of the two operations depends on the input state and squeezing regime, with their fidelity curves approaching one another in the ideal transmissivity limit. These findings establish that the benefits from non-Gaussian engineering depend on the specific operation, the nature of the input state, and the relevant parameter regime. To account for the probabilistic nature of resource preparation, we have additionally examined the probability-weighted fidelity enhancement \(P\cdot\Delta F\). This comparison shows that the relative advantage inferred from the fidelity enhancement alone can change when the success probability of the non-Gaussian operation is considered. In particular, the symmetric \((1,1)\) photon subtraction configuration provides the most favorable probability-weighted enhancement.
Moreover, accounting for the probabilistic preparation of the non-vacuum ancillary Fock states required for photon addition introduces an additional preparation penalty that is absent for vacuum-assisted photon subtraction.

In summary, our results establish a framework for evaluating the relative effectiveness of photon addition and photon subtraction in teleportation using an already non-Gaussian entangled resource. Within the configurations and parameter regimes studied, photon subtraction, particularly the symmetric \((1,1)\) pattern, offers the optimal trade-off between fidelity enhancement and preparation probability. These results also highlight the importance of considering both teleportation performance and resource-generation costs
when investigating the effectiveness of non-Gaussian operations in CV quantum teleportation.
\label{conclusion}
\appendix
\section{Mathematical details for PA-TMSFS teleportation protocol}
\label{App}
\subsection{Wigner characteristic functions of input states}
We employ the phase-space formalism developed earlier for the PS-TMSFS resource \cite{Ankita2026PLA}. The reader is referred to that work for a comprehensive derivation. The present section summarizes the essential definitions and expressions required for the fidelity calculations.

For an $n$-mode bosonic system, the quadrature operators
$\hat{\boldsymbol{\eta}}=(\hat{q}_1,\hat{p}_1,\dots,\hat{q}_n,\hat{p}_n)^T$ satisfy the canonical commutation relations
\begin{equation}
	[\hat{\eta}_i,\hat{\eta}_j]
	=
	i\Omega_{ij},
\end{equation}
where $\Omega=\bigoplus_{k=1}^{n}\omega$ is the symplectic matrix with
\begin{equation}
	\omega=
	\begin{pmatrix}
		0 & 1\\
		-1 & 0
	\end{pmatrix}.
\end{equation}
A state $\hat{\rho}$ is described in phase space by its Wigner characteristic function as
\begin{equation}
	\chi(\Lambda)
	=
	\mathrm{Tr}
	\left[
	\hat{\rho}\,
	e^{-i\Lambda^T\Omega\hat{\boldsymbol{\eta}}}
	\right],
\end{equation}
where $\Lambda\in\mathbb{R}^{2n}$. For a Gaussian state with displacement vector $\mathrm{d}=\langle\hat{\eta}\rangle$ and covariance matrix $V_{ij}=\tfrac{1}{2}\langle\{\Delta\hat{\eta}_i,\Delta\hat{\eta}_j\}\rangle$,
the WCF takes the form
\begin{equation}
	\chi(\Lambda)
	=
	\exp\!\left[
	-\frac{1}{2}\Lambda^{T}(\Omega V \Omega^T) \Lambda
	-i(\Omega\mathrm{d})^T \Lambda
	\right].
	\label{WCF_GS}
\end{equation}
Accordingly, the Wigner characteristic functions of coherent and squeezed vacuum states can be expressed as follows
\begin{align}
	\chi_{\mathrm{coh}}(\Lambda)
	&=
	\exp\!\left[
	-\frac{1}{4}(\tau^2+\sigma^2)
	-i(\tau d_p-\sigma d_q)
	\right],
	\label{WCF_CS}\\
	\chi_{\mathrm{squ}}(\Lambda)
	&=
	\exp\!\left[
	-\frac{1}{4}
	\left(
	\tau^2 e^{2r}
	+
	\sigma^2 e^{-2r}
	\right)
	\right].
	\label{WCF_SS}
\end{align}
For the non-Gaussian inputs considered here, the WCF of the single-mode Fock state $\ket{n}$ is
\begin{equation}
	\chi_{|n\rangle}(\tau,\sigma)
	=
	\exp\left[
	-\frac{\tau^2}{4}
	-\frac{\sigma^2}{4}
	\right]
	L_n
	\left(
	\frac{\tau^2+\sigma^2}{2}
	\right),
	\label{Lag}
\end{equation}
where $L_n(x)$ is the Laguerre polynomial of order $n$, that can be equivalently expressed via the generating function as
\begin{align}
	\chi_{|n\rangle}(\tau,\sigma)
	&	=
	\exp\left[
	-\frac{\tau^2}{4}
	-\frac{\sigma^2}{4}
	\right]
	\hat{F}
	e^{2st+s(\tau+i\sigma)-t(\tau-i\sigma)}, \\
	\text{with}	\qquad	&\hat{F}
	=
	\frac{1}{2^n n!}
	\frac{\partial^n}{\partial s^n}
	\frac{\partial^n}{\partial t^n}\{\bullet\}
	\Bigg|_{s=t=0}. \nonumber
	\label{WCF_Fock}
\end{align}
The even and odd Schr\"odinger cat states, $\ket{\psi_{\pm}}=\mathcal{N}_{\pm}\left(\ket{\gamma}\pm\ket{-\gamma}\right)$ with $\mathcal{N}_{\pm}=\left[2(1\pm e^{-2|\gamma|^2})\right]^{-1/2}$ have the characteristic functions as
\begin{equation}
	\chi_{\pm}(\alpha)
	=
	\frac{
		e^{-|\alpha|^{2}/2}
	}
	{1\pm e^{-2|\gamma|^{2}}}
	\left[
	\cosh
	\left(
	\alpha\gamma^{*}
	-
	\alpha^{*}\gamma
	\right)
	\pm
	e^{-2|\gamma|^{2}}
	\cosh
	\left(
	\alpha\gamma^{*}
	+
	\alpha^{*}\gamma
	\right)
	\right].
	\label{WCF_CAT}
\end{equation}
\subsection{Matrices for the PA-TMSFS characteristic function 
	}
	The two-mode squeezing operator $S_2(r)$, beam-splitter transformation $B_{ij}(T)$, and their action on the displacement vector, covariance matrix, and WCF are discussed in detail in \cite{Ankita2026PLA}. Avoiding unnecessary repetition, 
	the matrices involved in the WCF of the PA-TMSFS [see Eq.~\eqref{PAWCF}] are given explicitly as
	%
	\begin{align}
M_1 &= 	-\frac{1}{4{a}_0}	\begin{pmatrix}
	a_1 & 0 & -a_2 &0 \\
	0 & a_1 & 0 & a_2 \\
	-a_2 & 0 & a_1 & 0 \\
	0& a_2 & 0 & a_1
\end{pmatrix}  \label{M_1}\\[0.2cm]
\text{with} ~~~~~~ & {a}_0 =  \beta^2 -\alpha^2 t_1^2 t_2^2, \nonumber \\ & {a}_1 =  \beta^2 +\alpha^2 t_1^2 t_2^2, \\  &{a}_2 =  2\alpha\beta t_1 t_2.\nonumber\\\nonumber \\
\text{Here} ~~~~~~~ & t_1=\sqrt{T_1},\,t_2=\sqrt{T_2}, \nonumber \\ &r_1^2+t_1^2=1,\,\,r_2^2+t_2^2=1,  \\ & \alpha=\sinh{\xi}, \nonumber \\
& \beta=\cosh{\xi}\nonumber, 
\end{align}
\begin{equation}
M_2 = \frac{1}{a_0}
\begin{pmatrix}
	b_1 & ib_1 & b_2 & -ib_2 \\
	-b_1 & ib_1 & -b_2 & -ib_2 \\
	b_3 & -ib_3 & b_4 & ib_4 \\
	-b_3 &- ib_3 &- b_4 & ib_4 \\
	b_5 & ib_5 & b_6 & -ib_6 \\
	-b_5 & ib_5 & -b_6 & -ib_6 \\
	b_7 & -ib_7 & b_8 & ib_8 \\
	-b_7 & -ib_7 & -b_8 & ib_8
\end{pmatrix},
\label{M_2}
\end{equation}
where
\begin{equation}
\begin{aligned}
	& b_1 = \beta t_1, \quad && b_5 = \beta^2 r_1 , \\
	& b_2 = -\alpha  t_1^2 t_2, \quad && b_6 = -\alpha \beta r_1 t_2t_1, \\
	& b_3 = -\alpha t_1 t_2^2, \quad && b_7 = -\alpha \beta r_2 t_1t_2, \\
	& b_4 = \beta t_2, \quad && b_8 = \beta^2 r_2 ,
\end{aligned}
\end{equation}
\begin{equation}
M_3 = \frac{1}{a_0}
\begin{pmatrix}
	0 & c_1 & c_2 & 0 &  0 &c_3 & c_4 & 0  \\
	c_1 & 0 & 0 & c_2 & c_3 & 0 &0 & c_4 \\
	c_2 & 0 & 0 & c_5 &  c_6 & 0 & 0 & c_7 \\
	0 & c_2 &  c_5 & 0 & 0 & c_6 & c_7 & 0 \\
	0 & c_3 & c_6 & 0 &  0 &c_8 & c_9 & 0  \\
	c_3 & 0 & 0 & c_6 & c_8 & 0 &0 & c_9 \\
	c_4 & 0 & 0 & c_7 &  c_9 & 0 & 0 & c_{10} \\
	0 & c_4 &  c_7 & 0 & 0 & c_9 & c_{10} & 0 
\end{pmatrix},
\label{M_3}
\end{equation}
with
\begin{equation}
\begin{aligned}
	& c_1 = t_1^2, \quad && c_6 = \alpha  r_1  t_2^2t_1, \\
	& c_2 = -\alpha \beta(1-t_1^2 t_2^2), \quad && c_7 = \beta r_2 t_2, \\
	& c_3 = r_1 \beta t_1 , \quad && c_8 = \beta^2 r_1^2, \\
	& c_4 = \alpha r_2 t_1^2 t_2, \quad && c_9 = \alpha \beta r_1 r_2 t_1 t_2, \\
	& c_5 = t_2^2, \quad && c_{10} = \beta^2 r_2^2 	\end{aligned}
	\end{equation}
	\subsection{Auxiliary matrices and coefficients in the fidelity expressions}
	The matrices $M_4$ and $M_5$, appearing in the coherent and squeezed vacuum fidelity expressions in Eqs.~\eqref{FCS} and \eqref{FSS} respectively, are defined as 
	\begin{equation}
M_4 = \frac{1}{d_0}
\begin{pmatrix}
	0 & c_1 & d_2 & 0 &  0 &c_3 & d_4 & 0  \\
	c_1 & 0 & 0 & d_2 & c_3 & 0 &0 & d_4 \\
	d_2 & 0 & 0 & c_5 &  d_6 & 0 & 0 & c_7 \\
	0 & d_2 &  c_5 & 0 & 0 & d_6 & c_7 & 0 \\
	0 & c_3 & d_6 & 0 &  0 &c_8 & d_9 & 0  \\
	c_3 & 0 & 0 & d_6 & c_8 & 0 &0 & d_9 \\
	d_4 & 0 & 0 & c_7 &  d_9 & 0 & 0 & c_{10} \\
	0 & d_4 &  c_7 & 0 & 0 & d_9 & c_{10} & 0 
\end{pmatrix},
\label{M_4}
\end{equation}
where
\begin{equation}
\begin{aligned}
	&	c_1 = t_1^2, \quad && d_2 = \frac{t_1t_2(\beta-\alpha t_1t_2) + 2\beta C_2}
	{\beta+\alpha t_1t_2},\\
	& c_3 = r_1\beta t_1, \quad && d_4 = \beta r_2t_1,\\
	& c_5 = t_2^2, \quad && d_6 = \beta r_1t_2,  \\
	& c_7 = \beta r_2t_2, \quad && c_8 = \beta^2r_1^2, \\
	& d_9 = \beta^2r_1r_2, \quad && c_{10} = \beta^2r_2^2.
\end{aligned}
\end{equation}
\begin{equation}
M_5 = \frac{1}{e_0}
\begin{pmatrix}
	e_1 & e_2 & e_3 & e_4 & e_5 & e_6 & e_7 & e_8 \\
	e_2 & e_1 & e_8 & e_3 & e_6 & e_5 & e_8 & e_7 \\
	e_3 & e_4 & e_9 & e_{10} & e_{11} & e_{12} & e_{13} & e_{14} \\
	e_4 & e_3 & e_{10} & e_9 & e_{12} & e_{11} & e_{14} & e_{13} \\
	e_5 & e_6 & e_{11} & e_{12} & e_{15} & e_{16} & e_{17} & e_{18} \\
	e_6 & e_5 & e_{12} & e_{11} & e_{16} & e_{15} & e_{18} & e_{17} \\
	e_7 & e_8 & e_{13} & e_{14} & e_{17} & e_{18} & e_{19} & e_{20} \\
	e_8 & e_7 & e_{14} & e_{13} & e_{18} & e_{17} & e_{20} & e_{19}
\end{pmatrix} \label{M_5}
\end{equation} 
where
\begin{equation}
\begin{aligned}
	& e_0 = 2(a_1+\delta a_0),\,\,\, e_1 = -\gamma t_1^2, \\
	& e_2 = t_1^2\left(\frac{a_0}{a_3}+\delta\right),\,\,\, e_3 = \frac{1}{a_0}
	\left[
	\frac{t_1t_2}{2}(a_3+\delta a_0)
	-\alpha\beta(1-t_1^2t_2^2)e_0
	\right], \\
	& e_4 = \gamma t_1t_2,\,\,\, e_5 = -\gamma t_1\beta r_1, \\
	& e_6 = \beta r_1t_1
	\left(\frac{a_0}{a_3}+\delta\right), \,\,\, e_7 = \frac{r_2t_1}{a_0}
	\left[(\beta+2\alpha t_1t_2)e_0-\beta a_2\right], \\
	& e_8 = \beta r_2t_1, \,\,\, e_9 = -\gamma t_2^2, \\
	& e_{10} = t_2^2\left(\frac{a_0}{a_3}+\delta\right), \,\,\,
	e_{11} = \frac{r_1t_2}{a_0}
	\left[(\beta+2\alpha t_1t_2)e_0-\beta a_2\right], \\
	& e_{12} = \gamma\beta r_1t_2, \,\,\, e_{13} = -\gamma\beta r_2t_2, \\
	& e_{14} = \beta r_2t_2
	\left(\frac{a_0}{a_3}+\delta\right), \,\,\, e_{15} = -\gamma\beta^2r_2^2, \\
	& e_{16} = \beta^2r_1^2 
	\left(\frac{a_0}{a_3}+\delta\right), \,\,\,  
	e_{17} = \frac{\beta r_1r_2}{a_0}
	\left[(\beta+2\alpha t_1t_2)e_0-\beta a_2\right], \\
	& e_{18} = \gamma\beta^2r_1r_2, \,\,\,  
	e_{19} = -\gamma\beta^2r_2^2, \\
	& e_{20} = \beta^2r_1^2
	\left(\frac{a_0}{a_3}+\delta\right), \\
	\text{Here}~~~~~~~~~
	& \mu = \sinh{2\epsilon},\\
	& \delta = \cosh{2\epsilon}.
\end{aligned} 
\end{equation}
The quantities $z$, $R$ and $S$ appeared in Eqs.~\eqref{FFS} and \eqref{FCAT} are
\begin{align}
z=\frac{a_1-a_2}{a_0}+1,
\label{B-z}
\end{align}
and
\begin{align}
R&=u^T
\begin{bmatrix}
	-(b_1+b_2)\\
	b_1+b_2\\
	-(b_3+b_4)\\
	b_3+b_4\\
	-(b_5+b_6)\\
	b_5+b_6\\
	-(b_7+b_8)\\
	b_7+b_8
\end{bmatrix},
\qquad
S=u^T
\begin{bmatrix}
	b_1+b_2\\
	b_1+b_2\\
	-(b_3+b_4)\\
	-(b_3+b_4)\\
	b_5+b_6\\
	b_5+b_6\\
	-(b_7+b_8)\\
	-(b_7+b_8)
\end{bmatrix} \label{R_S}
\end{align}
\section*{Acknowledgements}	
Ankita's work is supported by the University Grants Commission (UGC), Government of India (Award No.~231610110670). A.~C. acknowledges Haryana State Council for Science, Innovation and Technology for the support provided through the Project No. HSCSIT/R$\&$D/2025/2060. 
\bibliography{paper2_ref}
\end{document}